\documentclass[journal,10pt,twocolumn,comsoc]{IEEEtran}

\usepackage[T1]{fontenc}

\ifCLASSINFOpdf
\else
\fi

\usepackage{lipsum}
\usepackage{setspace}
\usepackage{algcompatible}
\usepackage{algorithmicx}
\usepackage{algpseudocode}
\usepackage{algorithm}

\usepackage{mathtools,lipsum,cuted}
\usepackage{amsmath}     
\usepackage{amssymb}                       
\usepackage{mathrsfs}                          
\usepackage{upgreek}
\usepackage{bm}     
\renewcommand\eqref[1]{(\ref{#1})}
\usepackage{booktabs}
\usepackage{booktabs}

\usepackage{enumitem}
\usepackage{graphicx}
\usepackage{multicol}
\usepackage{caption}

\usepackage{mathtools,amssymb,lipsum}

\usepackage{cuted}
\usepackage{wrapfig}

\usepackage{srcltx}
\usepackage{amsmath}
\usepackage{amssymb}
\usepackage{cite}
\usepackage{psfrag}
\usepackage{epsfig}
\usepackage{subfigure}
\usepackage{amsthm}
\usepackage{thmtools}
\usepackage{color}
\usepackage{stfloats}
\usepackage{multirow}

\newtheorem{thm}{Theorem}

\begin{document}

\title{ \LARGE Error Correction Coding Meets Cyber-Physical Systems: \\
Message-Passing Analysis of Self-Healing Interdependent Networks}

\author{Ali Behfarnia, ~\IEEEmembership{Student Member,~IEEE,} Ali Eslami, ~\IEEEmembership{Member,~IEEE,}%

\thanks{The material in this paper was presented in part at the 25th International Conference on Computer Communication and Networks (ICCCN), and IEEE Globecom Workshops (GC Wkshps), both in 2016, \cite{Behfar_Eslami} and \cite{behfarnia2016dynamics}, respectively.}

\thanks{A. Behfarnia and A. Eslami are with the Department of Electrical Engineering and Computer Science, Wichita State University, Wichita, KS, USA ~(emails: axbehfarnia@shockers.wichita.edu, ali.eslami@wichita.edu).}
}
\IEEEtitleabstractindextext{%
\begin{abstract}
Coupling cyber and physical systems gives rise to numerous engineering challenges and opportunities. An important challenge is the contagion of failure from one system to another, which can lead to large-scale cascading failures. However, the \textit{self-healing} ability emerges as a valuable opportunity where the overlaying cyber network can cure failures in the underlying physical network. To capture both self-healing and contagion, this paper considers a graphical model representation of an interdependent cyber-physical system, in which nodes represent various cyber or physical functionalities, and edges capture the interactions between the nodes. A message-passing algorithm is proposed for this representation to study the dynamics of failure propagation and healing. By conducting a density evolution analysis for this algorithm, network reaction to initial disruptions is investigated. It is proved that as the number of message-passing iterations increases, the network reaches a steady-state condition that would be either a complete healing or a complete collapse. Then, a sufficient condition is derived to select the network parameters to guarantee the complete healing of the system. The result of the density evolution analysis is further employed to jointly optimize the design of cyber and physical networks for maximum resiliency. This analytical framework is then extended to the cases where propagation of failures in the physical network is faster than the healing responses of the cyber network. Such scenarios are of interest in many real-life applications such as smart grid.
Finally, extensive numerical results are presented to verify the analysis and investigate the impact of the network parameters on the resiliency of the network. 
\end{abstract}

}

\maketitle
\IEEEdisplaynontitleabstractindextext
\IEEEpeerreviewmaketitle


\section{Introduction}

\IEEEPARstart{A} cyber-physical system (CPS) is a system of collaborating computational elements controlling physical entities. The future smart grid, intelligent transportation systems, distributed robotics, and medical monitoring systems are all examples of CPSs.
The interconnected nature of such systems gives rise to numerous engineering challenges and opportunities.
An important challenge is the contagion of failure from one system to another in a coupled system. Such contagion may lead to large-scale catastrophic failure triggered by a small failure, such as the 2003 blackout in the Northeastern United States \cite{andersson2005causes}. Here, the \emph{self-healing} ability emerges as a valuable opportunity, where the overlaying cyber network can cure failures in the underlying physical network. For example, after detecting the failure in the power system, smart grid exploits self-healing abilities such as control of the production and distribution of electricity to halt the damage instantaneously \cite{amin2008challenges}. In another example, the traffic control network that monitors taxi transportation could avoid congestion by calculating the fastest routes during a given time of the day \cite{hull2006cartel}.

The study of interdependent networks was sparked by the seminal work of Buldyrev et al. \cite{Buldyrev10}, where a simple ``one-to-one" interdependence model was considered. Several authors \cite{buldyrev2011interdependent, gao2012networks, cho2010correlated, Boccaletti_Survey, cellai2013percolation, PhysRevE_Zanj, huang2011robustness, parshani2010interdependent, zhou2012assortativity, yagan2012optimal,schneider2013towards, shao2010cascade, huang2013balancing, huang2015small, gao2012robustness, sen14, luo14nature, shahrivar2015robustness, Expander-xheal11,Healing_PhysRevE15, Self-healingPloS14, liu2014modeling, EpidemicMIT15, Switching13, Healing_Elsevier14, majdandzic2016} then aimed to extend findings to more realistic scenarios (a brief overview is provided in Section \ref{sec:relatedwork}). However, self-healing, and its modeling and design advantages in cyber-physical systems have been mostly overlooked in the literature.
Among the most important issues in the design of future CPSs such as the smart grid are the following:
\begin{itemize}
\item Derive an analytically-tractable model that captures the key features of real-life systems such as self-healing and contagion,
\item Develop a framework that enables studying multiple layers of interconnected cyber and physical systems.
\end{itemize}

In this paper, we take a novel approach to address these issues by applying ideas inspired by error correction coding to model, analyze, and design cyber-physical systems. Our main contributions can be summarized as follows:
\begin{itemize}
\item We propose a graphical model representation of CPSs, where nodes represent network functionalities of the cyber and physical nodes, and the edges represent the connections within each network and between the two.
\color{black}
\item We apply a message-passing algorithm to the CPS graphical model, where messages represent the interactions between the nodes, including contagion and healing. We then borrow the concept of \emph{density evolution} used in performance analysis of codes on graphs such as repeat-accumulate (RA) codes \cite{jin2000irregular}, low-density parity-check (LDPC) codes \cite{shokrollahi, richardson2001capacity}, and turbo codes \cite{divsalar2001iterative}, to study the dynamics of failure propagation and healing in CPSs after an initial disturbance. The result is a closed form formula referred to as ``density evolution equation".
\item From the density evolution equation, we derive a sufficient condition for choosing the network parameters to guarantee complete healing of the system. This condition provides us with simple yet critical design guidelines.
\item Density evolution equation is then employed to set up an optimization problem to do the following: (a) find the most severe initial disruption that can be tolerated by the system, given the network parameters, and (b) find the optimal values of the network parameters for maximum resilience against an initial disruption.
\item We extend our analysis to study the behavior of a CPS in the presence of time delays in cyber nodes' response to physical failures. Such delays, common in many real-life systems, could be for a number of reasons, such as recovering data from the database, collecting data from other physical nodes, gathering information from neighboring cyber nodes, etc. We derive the density evolution equation in the presence of such delays.
\item We provide a steady state analysis showing that, as time goes by, the network reaches a steady state condition that would be either a complete healing or a complete collapse. 
\item We provide extensive numerical results to verify the analysis and investigate the impact of the network parameters on the resiliency of the network. The results confirm the steady state behavior, and the largest tolerable disturbance is obtained for many networks through solving the optimization problem. It is also observed that to preserve the healing ability of the system, the probability of failure propagation {\em among physical nodes} should be kept small. Another important observation clarifies that if the cyber nodes' delay takes more than a few time slots, the probability of complete healing will significantly reduce.
\end{itemize}

The rest of the paper is organized as follows. Section \ref{sec:relatedwork} provides a brief overview of the related work. \color{black} Section \ref{sec:BP} provides a brief introduction to the use of message passing and density evolution in graphical models, particularly factor graphs. \color{black} Section \ref{sec:Buldyrev} starts the analysis by applying a message-passing algorithm to a simple self-healing one-to-one network inspired by Buldyrev's model. Then, Section \ref{sec:formulation} describes the system model, notations, and message passing in the general CPS. Section \ref{sec:Fixed_Point_Analysis} provides a density evolution analysis of the proposed message-passing algorithm. This section also includes a sufficient condition for the system to be healed, optimizing network parameters for maximum resiliency, and the effect of processing time delay in the network. Section \ref{sec:results} is devoted to extensive numerical results, and Section \ref{sec:conclusion} concludes the paper.

\section{Related Work}\label{sec:relatedwork}
Several papers extended the findings of Buldyrev et al. \cite{Buldyrev10} by applying percolation theory while focusing on the size of the remaining giant component after a cascading failure. Authors in\cite{buldyrev2011interdependent} studied the percolation of failures after an attack in a one-to-one interdependent network model in which mutually dependent nodes have the same number of neighbors. 
\color{black} Parshani et al. \cite{parshani2010interdependent} studied the case where only a fraction of the nodes in both networks depend on each other, that is, some nodes in each network are not connected to the other network. They proved that the reduction of coupling between networks leads to a change from a first-order percolation phase to a second-order percolation phase. Later, a systematic strategy based on betweenness was introduced \cite{schneider2013towards} to select a minimum number of autonomous nodes that guarantees a smooth transition. This reduces the fragility of the network without losing functionality. Shao et al. \cite{shao2010cascade} proposed an interdependent model, taking into account the realistic scenarios at which a node in a network $X$ might be supported by more than one node in a network $Y$, and vice versa. In such cases, a node will continue to work as long as at least one of its supporting nodes is still working. \color{black} Gao et al. \cite{gao2012robustness} have developed an analytical framework for studying the robustness of tree-like $n$ interdependent networks. They found that for any $n \geq 2$ (for Erd$\ddot{\text{o}}$s-R$\acute{\text{e}}$nyi, random regular, and scale-free graphs), cascading failure appears, and transition becomes a first-order compared to a second-order transition.  
\color{black}
 A ``regular allocation" algorithm was proposed in \cite{yagan2012optimal} to allocate the same number of interlinks to each node. Authors proved that such allocation is optimal for a network with an unknown topology, and that employing bidirectional interlinks instead of unidirectional ones leads to better robustness. In a different line of work, several authors \cite{huang2013balancing}, \cite{huang2015small} studied the influence of active small clusters appearing after an attack on the whole network performance. In particular, they obtained an upper bound for the fraction of operating active small clusters after a cascading failure.
\color{black} Shahrivar et al. \cite{shahrivar2015robustness} studied the resilience of random interdependent networks through algebraic connectivity. They obtained a threshold for $r$-robustness, which is the same as that required for the graph to have a minimum degree $r$. 

\color{black}
A number of works \cite{Expander-xheal11,Healing_PhysRevE15, Self-healingPloS14, liu2014modeling, EpidemicMIT15, Switching13} have been devoted to self-healing single-layer networks. Several authors \cite{Self-healingPloS14, Healing_PhysRevE15} studied the concept of self-healing networks through distributed communication protocols that set up new links to recover the system connection. In particular, Quattrociocchi et al. \cite{Self-healingPloS14} evaluated the performance of redundant links in small-world networks against link failures. Through small-world topologies, they found that some long-range connections could greatly increase the resiliency of the network. Liu et al. \cite{liu2014modeling} investigated the effect of restoration time and resources to study the cascading overload failure in homogeneous (e.g., Erd$\ddot{\text{o}}$s-R$\acute{\text{e}}$nyi) and heterogeneous (e.g., scale-free) networks. For an SIS-type epidemic process, Drakoloulos et al. \cite{EpidemicMIT15} achieved a lower bound on the optimal expected extinction time. This bound was obtained as a function of curing budget, maximum degree of each node, and epidemic parameters. Besides, many real-time technologies in realistic networks that support the realization of self-healing methods have been developed. For example, Li et al. \cite{Switching13}   
developed an optimization problem for a protection strategy (e.g., switching transmission line) in a critical infrastructure (e.g., power grids). 

A few works \cite{Healing_Elsevier14, majdandzic2016} have recently studied self-healing multi-layer networks. Stippinger and KertÃl'sz \cite{Healing_Elsevier14} introduced a healing strategy for an interdependent network based on the formation of new links. By applying recovering links after failures, they found that the increase in resiliency of an interdependent network has power-law scaling with the probability of healing. They also showed that it is possible to suppress the cascading failure by keeping the healing probability above a critical value. Majdandzic et al. \cite{majdandzic2016} developed a phase diagram for multi-layer networks to find an optimal repairing strategy in damaged interacting systems. 

Our work is concerned with the study of failure, the evolution of failure, and the recovery process in a cyber-physical network. The main differences between this paper and the above literature are as follows:  i) we apply, for the first time, a message-passing analysis to study the dynamics of failure propagation and healing in a cyber-physical system, ii) we obtain a closed-form equation for the evolution of failure in a self-healing interdependent network, iii) we develop a closed-form equation for the evolution of failure in the presence of time delay for recovering nodes in a self-healing interdependent network, and iv) we derive a sufficient condition for a self-healing cyber-physical system that prevents cascading failure in the network. In all the above, we exploit techniques from coding theory to analyze network resiliency. 
 \vspace{-.2 in}
\section{Message Passing and Density Evolution in Graphical Models}\label{sec:BP}
This section provides a background on message passing in factor graphs, which is required to understand the analysis in this paper.
We start with a brief introduction to factor graphs and their message passing. 
We then explain message passing in the simple example of LDPC decoding, followed by a description of density evolution as it is done in codes on graphs. Finally, we present the similarities and differences between codes on graphs and CPSs in their factor graphs, message passing, and density evolution.
\vspace{-4.5 mm}

\subsection{Factor Graphs}
A factor graph could be defined as a bipartite graph that expresses the factorization of a function. To explain this, let ${g(x_1,x_2,...,x_n)}$ be a function that could be factored into product of several local functions, each having a subset of all variables, ${\{x_1, x_2, ...,x_n\}}$. So, we have
\begin{align}
g(x_1,x_2,...,x_n) = \prod_{k \in K} f_{k}(X_{k})
\end{align}
where $K$ is a discrete index set, $X_k$ is a subset of ${\{x_1, x_2, ...,x_n\}}$, and $f_{k}(X_{k})$ is a function with the elements of $X_k$ as arguments.  A \emph{factor node} denotes a local function $f_k$, a \emph{variable node} denotes each variable $x_k$, and an edge exists between variable node $x_k$ and factor node $f_k$ if and only if $x_k$ is an argument of $f_k$. Fig. \ref{fig:Factor_graph} shows an example of a function, $g(x_1,x_2,x_3,x_4,x_5)$, on a bipartite graph that can be obtained as  the product of $f_1(x_1,x_2) f_2(x_2) f_3(x_1,x_3) f_4(x_2,x_4) f_5(x_3,x_5)$.

A popular message-passing algorithm on factor graphs is the sum-product message-passing algorithm, also known as belief propagation, which computes all marginals of the individual variables of the function. Computation of a marginal function begins with the leaves of a factor graph. Each leaf variable node and leaf factor node send ``belief'' messages to their parents. A variable node simply sends the product of all received messages as its belief, while a factor node with parent $x$ calculates the product of all received messages from its children, and then operates on the result with a sum over all its variables except $x$, $\sum_{\sim\{x\}}$, to send its belief. It is worth noting that the role of parents and children nodes are temporary, and it would be variant for different marginalized parameters. 
\begin{figure}[t]
\centering
{\includegraphics[width = 2.5 in , height = 1.2 in]{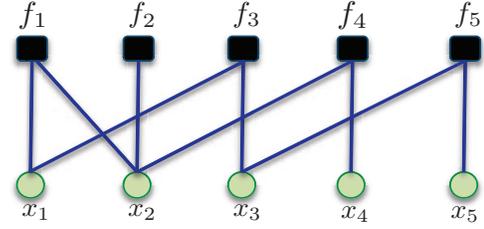}}
\captionof{figure}{\color{black} A factor graph for the product $f_1(x_1,x_2) f_2(x_2) f_3(x_1,x_3) f_4(x_2,x_4) f_5(x_3,x_5)$.}\label{fig:Factor_graph}
\end{figure}\label{fig:Factor_graph}
\begin{figure}
\centering
 {\includegraphics[width =3.2 in , height=1.7 in]{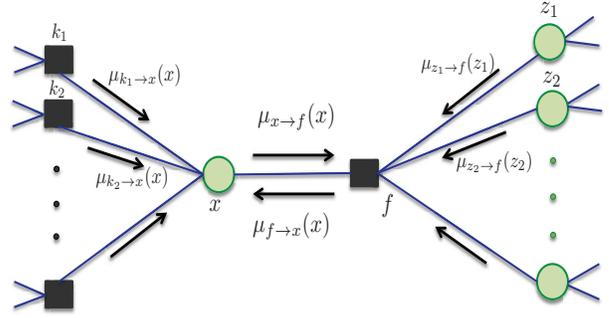}}
\captionof{figure}{\color{black} A part of factor graph, showing the update rules of sum-product message-passing algorithm.}
 \end{figure}\label{fig:Factor_messages}\color{black}
To obtain a mathematical expression for the message-passing algorithm, let $n(y)$ denote a set of all neighbors of a node $y$, $\mu_{f\rightarrow x}(x)$ represents a message sent from a factor node, $f$, to a variable node, $x$, and $\mu_{x \rightarrow f}(x)$ shows the message sent from node $x$ to $f$. Then, as illustrated in Fig. \ref{fig:Factor_messages}, the sum-product message-passing algorithm can be written as follows \cite{Kschischang2001}:
\begin{align}
\mu_{\, x\rightarrow f}(x) &=  \prod_{k \, \in \, n(x) \setminus \{f\}} \mu_{\, k\rightarrow x}(x) \\
\mu_{\, f\rightarrow x}(x) &= \sum_{\sim\{x\}} \bigg( f(X)    \prod_{z \, \in \, n(f) \setminus \{x\}} \mu_{\, z\rightarrow f}(z)  \bigg).
\end{align}
\color{black}For example, $g(x_4)$ can be obtained as follows:
\begin{align}
g(x_4) = \mu_{f_4 \rightarrow x_4}(x_4)
\end{align}
where,
\begin{align}
\mu_{f_4 \rightarrow x_4}(x_4) = \sum_{\{x_2\}} f_4(x_2,x_4) \: \mu_{x_2 \rightarrow f_4}(x_4)\\
\mu_{x_2 \rightarrow f_4}(x_4) = \mu_{f_2 \rightarrow x_2}(x_2) \: \mu_{f_1 \rightarrow x_2}(x_2).
\end{align}
%

\subsection{Message Passing in Codes on Graphs} 
\begin{figure}[t]
\centering
{\includegraphics[width =3.4 in , height=1.5 in]{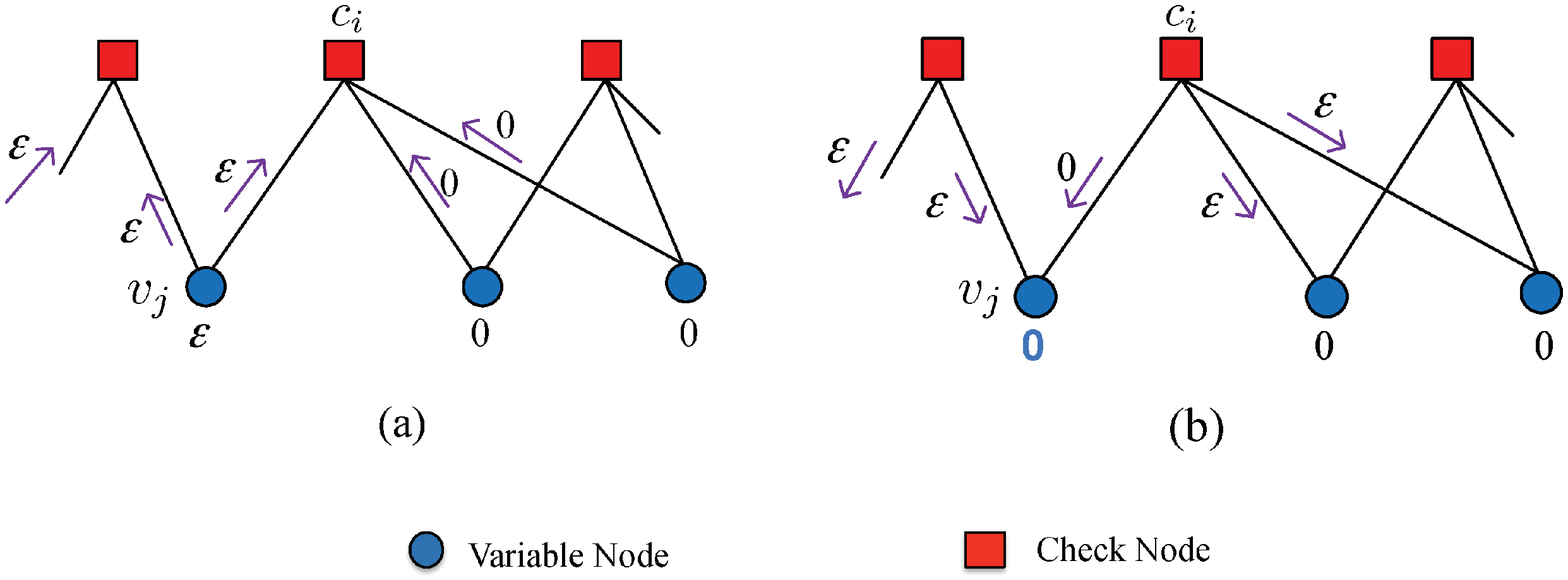}}
\caption{One iteration of message passing in LDPC
codes over binary erasure channel.}
\label{fig:BP_LDPC}
\end{figure}

Message passing has been successfully employed in the decoding of codes on graphs. 
Here we explain the message-passing decoding of LDPC codes with a simple example. Fig. \ref{fig:BP_LDPC} shows part of the \emph{Tanner graph} of an LDPC code, where the circles and squares represent, respectively, \emph{variable nodes} and \emph{check nodes}. The variable nodes correspond to the symbols received from the channel, i.e., channel outputs. In this particular example, we assumed a \emph{binary erasure channel} (BEC) where the channel outputs are either received correctly or are unknown. In a BEC, there is no bit flip from 0 to 1, or vice versa. The functionality of a check node is to do a check-sum on the values of its variable nodes and ensure that they add up (in modulo 2) to zero.

The goal of the decoder is to determine the correct values of the unknown bits after  multiple rounds of message passing between the variable and check nodes.  As shown in Fig. \ref{fig:BP_LDPC}, there is one unknown bit (variable node) at the channel output. At the beginning of each iteration, every variable node $v_j$ sends its value to all of its neighboring check nodes, as shown in Fig. \ref{fig:BP_LDPC} (a).
Every check node $c_i$ then derives what it believes about the value of each of its neighboring variable nodes, and sends the information back to each of them as a message. To derive this value for $v_j$, $c_i$ uses the messages received from all of its variable nodes, excluding $v_j$. If one or more of these messages are $\epsilon$ (erasure), then $c_i$ cannot be of help to $v_j$ at this round and therefore sends an $\epsilon$ to $v_j$. Otherwise, if all of these messages are either 0 or 1, then $c_i$ takes their check-sum and sends the result to $v_j$ as what it believes $v_j$ should be. An example of messages sent from check nodes to variable nodes is shown in Fig. \ref{fig:BP_LDPC}(b). The last two steps are repeated until the values of all variable nodes are derived, or a certain number of iterations is reached.
\subsection{Density Evolution in Codes on Graphs}
For the message-passing decoding of codes on graphs on general memoryless channels, messages between variable nodes and check nodes are often defined as log-likelihood ratios (LLRs) of probabilities that a given bit is ``1'' or ``0''. Since LLRs are often continuous variables, the probability of a message for a specific value of LLR can be described by a probability density function (pdf). Tracking the evolution of this pdf in a message-passing decoder is called \emph{density evolution} (DE) and can reveal the performance of the decoder. While DE is typically used for channels like binary additive white gaussian noise channel (BIAWGNC) with continuous LLRs,  this term can also be employed to study the evolution of erasures in BEC channels where LLRs are discrete. In this case, DE keeps track of the density of erasure messages ($\epsilon$) to analyze the performance of the decoding algorithm. For this, let $p_i$ denote the probability of $\epsilon$ message from a variable node to a check node, and let $q_i$ denote the probability of $\epsilon$ from a check node to a variable node, both in the \emph{i}-th iteration of message passing in Fig. \ref{fig:BP_LDPC}. The probability of $\epsilon$ message on the (\emph{i+1})-th iteration of message passing from a variable node with degree $d$, say $v_j$,  to a check node, say $c_t$, can be written as $p_{i+1} = p_{i}\, q_i^{d-1}$ under the independence assumption \cite{shokrollahi}. This equation is actually obtained using eq. (2). The term $q_i^{d-1}$ accounts for all received messages to $v_j$ except $c_t$ (to avoid positive feedback) which is going to receive a message $ \prod_{k \, \in \, n(x) \setminus \{f\}} \mu_{\, k\rightarrow x}(x) $. It was proved in \cite{shokrollahi} that the decoding algorithm is successful if the inequality $p_{i+1} < p_i$ holds for every $i\geq 0$. 

The recursive equation above is referred to as ``density evolution equation", and was obtained  for message passing in LDPC codes over a BEC. Similar equations can be derived for message-passing algorithm over other channels and for other types of codes on graphs (see \cite{jin2000irregular, richardson2001capacity, divsalar2001iterative}). In this paper, we apply the density evolution analysis to the factor graph of CPSs, and derive the density evolution equation for them.
\subsection{Codes on Graphs vs. Cyber-Physical Systems} 
\color{black} There are a few similarities and differences between codes on graphs and cyber-physical systems in the structure of their factor graphs and properties of messages. In codes, factor nodes check/correct variable nodes; in CPSs, cyber nodes can {\em heal} physical nodes. Similar to the codes, there are two types of nodes in a CPS, and their interactions can be captured through messages exchanged between them. However, some differences between the two applications can be recognized as follows.
\begin{enumerate}
\item In codes on graphs, unknown (damaged) variable nodes cannot affect the functionality of the check nodes. In other words, \emph{failure} cannot propagate from variable nodes to check nodes. On the other hand, in cyber-physical systems, failure of the physical components could cause a failure in the cyber network, and vice versa \cite{Buldyrev10}.
\item Factor nodes in a CPS factor graph may assume a more complicated functionality than in codes on graphs. Also, the delivery of the messages in a CPS may not be guaranteed. These result in a different message structure and, probably, a more complicated density evolution equation.
\item The Tanner graph of a code is a \text{bipartite} graph in which every edge connects a check node to a variable node. In other words, there are no edges connecting the variable nodes or check nodes. In a CPS, however, both of the cyber and physical systems are connected networks. Hence, a physical (cyber) node can directly affect the operation of other physical (cyber) nodes. 
\end{enumerate}
\color{black}

In this paper, we show how message passing can be applied to CPSs despite these differences.
The first two differences can be addressed by appropriately defining the messages, and physical and cyber node functions, while accounting for imperfect message passing. This will be presented in Sections V and VI. The last difference can be addressed using a rather standard approach, i.e., by adding virtual check  nodes to the CPS factor graph. This will be presented in details in the proof of Theorem 2 in Section VI.  
To provide intuition into our approach, we start by considering a slightly modified version of Buldyrev et al.'s ``one-to-one" network \cite{Buldyrev10}. We will then extend our analysis to more general cases of CPSs and include cyber-node time delays.

\color{black}
\section{Message Passing over a Self-Healing One-to-One Model}\label{sec:Buldyrev}
\begin{figure}[t]
\begin{center}
\centering
    \includegraphics[width=0.45\textwidth]{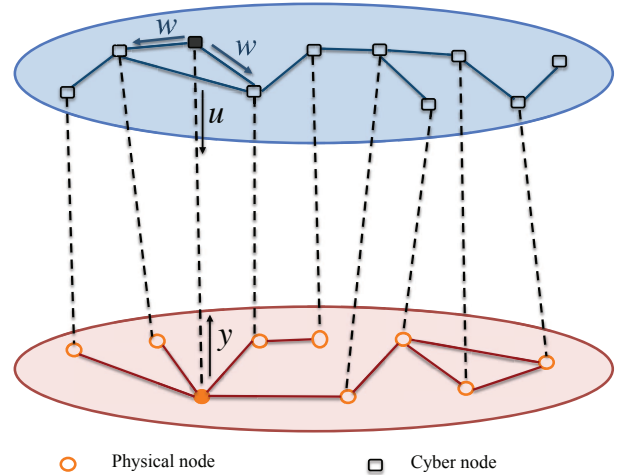}
\end{center}
\caption{Illustration of ``One-to-one" interdependent model.}\label{fig:One_2_One}
\end{figure}

Buldyrev et al. \cite{Buldyrev10} introduced a simple ``one-to-one" model that yields important insight into studying interdependent networks. In this model, it is assumed that two networks, say A and B, have the same number of nodes, N. The state (failed or alive) of a node in network A directly depends on the state of the

corresponding node in network B. Fig. \ref{fig:One_2_One} shows such a one-to-one model of a cyber-physical network. There will be an initial attack on the physical network failing each physical node with a probability $\epsilon$. Failures then propagate, not through the physical network but from the physical nodes to the cyber nodes and then through the cyber network. A cyber node with only failed cyber neighbors will fail, hence failing its underlying physical node.
As time passes and failure propagates between the two networks after several \emph{iterations}, a catastrophic cascade of failures may occur. 

In the model of Buldyrev et al., if a physical node fails, then the corresponding cyber node will also be lost, and there is no healing capability for either physical or cyber nodes. We slightly modify this model to consider a healing ability for cyber nodes. We assume that a cyber node that is not isolated from the cyber network can heal its failed physical node. That is, a cyber node with at least one healthy cyber neighbor still has access to the cyber network's data and can heal its physical node.

We capture the propagation of failure and healing between the nodes as \emph{defection} (D) and \emph{healing} (H) \emph{messages} exchanged between them, and apply message-passing analysis tools to study the evolution of the cascade in this interdependent network. We may look at this evolution within one (any) iteration, and see how the failure probability changes for physical nodes. If this probability increases at the end of the iteration, then a cascade will occur, and if it decreases at the end the iteration, then the network will heal completely. Let us consider the first iteration after the initial attack. Each physical node is failed at the beginning with probability $\epsilon$. Let $y$ denote the probability of a D message from a physical node to its cyber neighbor. Thus,
\begin{align}
y= \epsilon. 
\end{align}
A cyber node with a failed physical node sends D messages to all its cyber neighbors reporting that it has lost its physical connection. Denote the probability of this event by $w$. Since each cyber node is connected to only one physical node,
\begin{align}
w=y=\epsilon.
\end{align}
A cyber node with only failed cyber neighbors sends a D message to its physical node; otherwise, it sends an H message healing the physical node. If we denote the probability of the former (sending a D message to the physical node) by $u$, then
\begin{align}
& u = \rho(w)=\rho(\epsilon), 
\text{where} \quad
 \rho(z)= \sum_{i\geq1} \rho_i  z^i
\end{align}\label{eq:x2}
is the degree distribution of the cyber nodes, and $\sum_{i\geq1} \rho_i =1$.  Here, $\rho_i$ is the fraction of cyber nodes with $i$ cyber neighbors. 
Let us denote the probability of a physical node failure by $x$: 
\begin{align}
x &=  \text{Pr} \: \bigg\{\text{Receiving a D message from the cyber node} \bigg\}
\end{align}
Also, assume that $x_l$ represents the $l$-th iteration of message passing. Now, we are at the end of the first iteration and $x_1$ can be written as: 
\begin{align}\label{eq:X_01}
x_1 = u = \rho(\epsilon). 
\end{align}
\vspace{-.2 in}
Note that for $0\leq\epsilon<1$,

\begin{align}
x_1 =\rho(\epsilon)= \sum_{i\geq1} \rho_i  \epsilon^i < \sum_{i\geq1} \rho_i  \epsilon= \epsilon.
\end{align}
Therefore, at the end of the first iteration, the probability of failure for a physical node decreases. The same analysis as above can be carried out for any iteration $l$. If we denote the  physical node failures at the beginning of iterations $l$ and $l+1$ by $x_l$ and $x_{l+1}$, respectively, then
\begin{align}
x_{l+1} < x_l,
\end{align}
which means that our simple (and somewhat intuitive) healing rule for the Buldyrev et al. network will always lead to its complete healing. This of course will not be the case for more complicated network models with complex contagion and healing rules. However, the message-passing approach used in this section can be generalized to develop a framework for studying such cases. The rest of this paper is dedicated to this task. Section \ref{sec:formulation} sets up the network model and formulates the message-passing problem for the general case. Section \ref{sec:Fixed_Point_Analysis} then generalizes the technique used here by applying a \emph{density evolution analysis} to study the dynamics of the cascade in the network.

\section{Problem Formulation and Modeling}\label{sec:formulation}
This section presents a graphical model for studying message passing in cyber-physical systems. First, we explain our network model for both physical and cyber networks. Then, we describe our models for the initial disturbance, healing, and contagion within each network and between the two networks. \color{black} Finally, we explain how our modeling framework can be applied, in an abstract level, to a network of autonomous cars as an example of cyber-physical systems.
\color{black}

\subsection{Network Model}
For our analysis, we consider random networks with given degree distributions as models of cyber and physical networks. This enables us to model random networks with arbitrary degree distributions such as scale-free networks with a power law degree distribution \cite{barabasi1999emergence}, and Erd\H{o}s-R\'enyi random graphs with a Bernoulli degree distribution \cite{bollobas1998random}. We define cyber (physical) degree of a node as the number of nodes in the cyber (physical) network connected to the node. In a similar fashion to codes on graphs, we use polynomials to represent the degree distributions of the networks:
\begin{align}
\rho(z)= \sum_{i\geq1} \rho_i  z^i, \text{and} \quad
\lambda(z)=\sum_{i\geq1} \lambda_i z^i
\end{align}
denote the degree distributions of the cyber and physical networks, respectively, where $\rho_i$ is the fraction of cyber nodes with cyber degree $i$, and $\lambda_i$ is the fraction of physical nodes with physical degree $i$.

To capture the interconnections between the two networks, two more polynomials are needed: one for the physical degree distribution of cyber nodes, and one for the cyber degree distribution of physical nodes. However, in order to simplify the presentation of results,  we assume that each cyber node controls $a$ physical nodes, while each physical node is connected to one cyber node. The analysis for the general case of degree distributions could be carried out along the same lines as the analysis in this paper.

\subsection{Initial Disturbance, Contagion, and Healing}\label{sec:setting}
Here, we explain our models for the initial disturbance, contagion within each system and between the two, and healing of the physical system by the cyber system. Our methodology, however, could be extended to a wide range of models.

\begin{itemize}[leftmargin=.19 in]
\item \textbf{Initial disturbance}: We assume that each physical node initially fails with a small probability $\epsilon$, where $\epsilon \ll 1$, \color{black} independently from other nodes\color{black}. In this paper, we only consider initial disturbance for the physical network. The analysis for the case of a cyber attack could be conducted in a similar fashion.

\item \textbf{Contagion within physical network}: After being defected, a physical node may defect each of its neighbors with probability $p$. This probabilistic model is commonly used in the literature for a range of applications \cite{Dobson04}.

\item \textbf{Healing of physical nodes}: A cyber node heals a physical node if that physical node is its only defected physical neighbor. An example of this could be a control center that has all measurements but one from the power grid, so it must derive the phase or voltage value for the remaining component.

\item \textbf{Contagion from physical to cyber system}: A cyber node with no functioning physical neighbor will go out of service. An example could be an internet server that looses its power supply in a power outage.

\item \textbf{Contagion within cyber system}: If all cyber neighbors of a cyber node are out of service, then the cyber node itself will go out of service. An example could be an internet server whose neighboring servers have all been disconnected from the network.
\end{itemize}

\subsection{Message Passing in Cyber-Physical Systems} \label{sec:Message_Types}
In our model, the interactions between nodes are represented by messages. Accordingly, all sorts of contagions and the healing process scenarios explained above could be interpreted in a message-passing framework as follows:

\begin{enumerate}
\item Defection (D) message:
\begin{itemize}
\item A defected physical node sends a defection message D to its cyber neighbors with the probability of $y$.
It also sends a message D to each of its physical neighbors with probability $p$.
\item A defected cyber node sends a D-message to its cyber and physical neighbors with the probability of $w$.
\item A functioning cyber node that cannot heal a physical node sends a D-message to that node with the probability of $w$.
\end{itemize}
\item Healing (H) message: A  cyber node that is able to heal a physical node sends a healing message H to that node.
\begin{figure}[t]
\centering
 {\includegraphics[width =3.0 in , height=2.4 in]{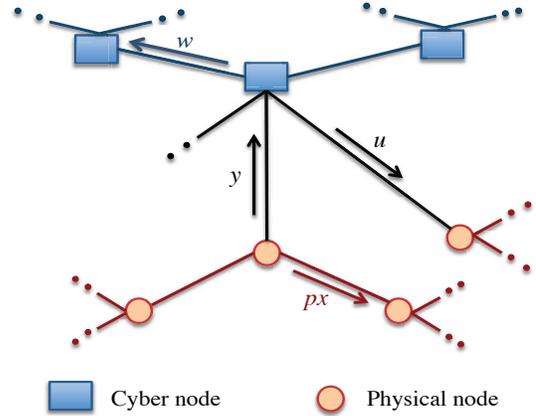}}
 \captionof{figure}{\protect\raggedright Example of messages exchanged in cyber-physical system.}\label{fig:Message}\end{figure}
\end{enumerate}
Defining the messages as above, shown in Fig. \ref{fig:Message}, addresses the first two differences listed in Section \ref{sec:BP} between codes on graphs and CPSs. Note that these messages are introduced to capture the interactions in the CPS, while they may not be actually exchanged between the nodes in the underlying networks.
\color{black} On the other hand, it is also possible that the messages do not arrive at the destinations due to imperfections of lines/channels in a real CPS. To address this point, we consider ``missing probabilities" for the messages. The probability of missing messages exchanged within the physical network, within the cyber network, and between the physical and cyber networks is represented by $P_{mp}$, $P_{mc}$, and $P_{mi}$, respectively. The value of these parameters may vary between 0 and 1 according to the underlying application.

\subsection{Autonomous Cars}
Recent advances in signal processing, communications, network monitoring, and control systems pave the way to the next generation of vehicles, such as autonomous cars. 
Autonomous cars have attracted investments from many companies such as Wayco (formerly known as the Google self-driving car), motivated by improved pollution control, ease of use, and safety. A self-driving car requires five basic functions in order to drive autonomously: localization, perception, planning, vehicle control, and system management \cite{jo2014development, jo2015development}. The localization function finds the estimated position of the vehicle based on GPS, and the perception function obtains the information of the surrounding environment using car sensors. Through this gathered information, the planning function provides the maneuvers of the self-driving car. The vehicle control function applies commands of the planning function by accelerating, braking, and steering the car. Finally, the system management function provides the supervision of self-driving cars. While the first four functions operate the physical aspects of the driverless car, the system management function forms a new network layer, called the \emph{vehicular cloud}, which maintains smooth traffic flow on roads through effective communication and distributed processing \cite{gerla2014internet}. Hence, a CPS could be defined with the vehicular cloud as the cyber network and autonomous cars as the physical network.
The message-passing framework introduced earlier in this section could then be applied to study the operation of this CPS as follows:
\begin{itemize}

\item \textbf{Initial disturbance}: Each autonomous car could fail due to any abnormality in the vehicle operation, from an engine problem to an unexpected speed or direction, as shown in Fig. \ref{fig:PPt_Auto}.

\item \textbf{Contagion within physical network}: Failure in an autonomous car could cause problems for neighboring cars. The failure could be healed by the supervising cyber nodes or it could result in cascading failure in the network \cite{dresner2008mitigating}.

\item \textbf{Healing of physical nodes}: Supervisors in the cyber network have access to essential data and services, from typical autonomous car measurements in the region to routes to service centers. The cyber node could use these services to help the autonomous car by the following: i) reporting a command to the planning function of the autonomous car in order to update the vehicle control or sending it to the nearest service center, or ii) providing an alarm to the passengers of the car to take appropriate action against the failure.

\item \textbf{Contagion from physical to cyber system}: If some physical nodes send the wrong measurements of their status on roads, then the supervising cyber node may improperly perceive the situation. 

\item \textbf{Contagion within cyber system}: False information in a cyber node would lead to sending wrong information to other cyber nodes. 

\end{itemize}

\begin{figure}[t]
\centering
 {\includegraphics[width =3.0 in , height=2.0 in]{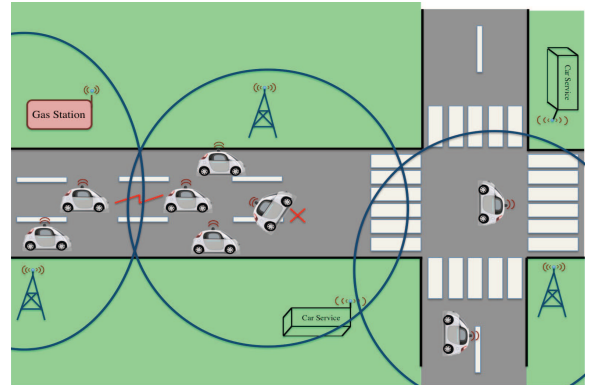}}
\captionof{figure}{\color{black}  Illustration of autonomous cars' message- passing to avoid congestion and failure propagation.}\label{fig:PPt_Auto}
\end{figure}\label{fig:PPt_Auto}

\section{Density Evolution Analysis of Cyber-Physical Systems}\label{sec:Fixed_Point_Analysis}
In order to study the impact of an initial disruption, we keep track of D and H messages by employing density evolution. Recall the example of Fig. \ref{fig:BP_LDPC} where variable and check nodes exchanged messages with values 0, 1, or $\epsilon$ in consecutive iterations. Density evolution tracks the density of D messages (e.g., $\epsilon$) as the number of iterations grows. This density is defined as the fraction of D messages among all messages exchanged between the variable nodes and check nodes at each iteration. The message-passing algorithm of Section \ref{sec:BP} is able to fix all damaged variable nodes, if and only if the density of D messages converges to 0 as  the number of iterations grows. Employing the concept of density evolution used in codes on graphs, we obtain a DE equation for CPSs.

\begin{thm} \label{th:DE}
The density evolution equation for the system defined in Section \ref{sec:formulation} can be obtained as follows:
\begin{flalign}\label{eq:f_l1}
\notag  x_0 &=  \epsilon, \\
 x_l &=  f(x_{l-1}), 
 \end{flalign}
where 
\begin{flalign}\label{eq:f_l2}
&f(x_{l-1}) = A \times B + A \times [1 - B ] \times P_{mi},
\end{flalign}
and
\begin{flalign}
\notag A &= x_{l-1} + \Big(1-x_{l-1}\Big)\Big(1-\lambda(1 - p x_{l-1})\Big) \Big(1 - P_{mp} \Big),
\end{flalign}
\begin{flalign}\label{eq:f_l2}
\notag B &= 1 - \Bigg\{  \Bigg[ \bigg( 1 -  x_{l-1} - \Big(1-x_{l-1}\Big)\Big(1-\lambda(1 - p x_{l-1})\Big) \\
\notag &\ \ \ \ \Big(1 - P_{mp} \Big) \bigg) \bigg(1 - P_{mi} \bigg)      \Bigg]^{a-1} \times \Bigg[1-\rho\bigg(   \Big(  \Big[ 1 - P_{mi} \Big] \\
\notag & \ \ \ \ \Big[  x_{l-1} + \Big(1-x_{l-1}\Big)\Big(1-\lambda(1 - p x_{l-1})\Big) \Big(1 - P_{mp} \Big) \Big]  \Big)^{a}  \\
\notag & \ \ \ \ \Big(1-P_{mc}\Big)\bigg) \Bigg] \Bigg\},
\end{flalign}
and $a$ is the number of physical nodes under each cyber node. The system heals if and only if $x_l\rightarrow 0$ as $l\rightarrow\infty$. 
\end{thm}
\color{black}

The proof for this theorem as well as others is presented in the appendix. The above formulas were obtained with the assumption of independence between messages in the network. In the following theorem, we utilize the LDPC code analysis \cite{richardson2001capacity} to show the existence of such an independence between messages in a CPS. 

\begin{thm}\label{th:Cycle_free}
For the cyber-physical system described in Section \ref{sec:formulation}, if the number of nodes is sufficiently large, then the incoming messages to each cyber or physical node can be considered as independent messages.
\end{thm}

\subsection{Steady-State Behavior of Cyber-Physical Systems}
We now study the steady-state behavior of the cyber-physical system against a failure for the defined message-passing rules. 
After a failure occurs in the network, defection messages appear in the network. It is expected that if the number of message-passing iterations increases, the density of defection messages through density evolution analysis approaches to $0$ or $1$, which means that the system reaches a steady-state condition that would be either complete healing or complete collapse. 
We prove this claim in the following theorem and confirm it via simulations in Section \ref{sec:results}.

\begin{thm}\label{th:Steady_State}
For the cyber-physical system defined in Section \ref{sec:formulation}, if the message-passing iterations increase, then the system will reach a steady-state condition, which is a complete-healing state or a complete-failure state. 
\end{thm}

\subsection{Sufficient Condition for Healing}
Once the recursive equation of density evolution is derived for a given set of contagion and healing rules, it can be utilized in many ways to gain useful insights into the network design. The following theorem, for example, employs equation (\ref{eq:f_l1}), \color{black}with the assumption of no missing messages ($P_{mi} = P_{mc} = P_{mp} = 0$), \color{black} to obtain a sufficient condition for the system to heal completely.

\begin{thm}\label{th:nec}
The cyber-physical system described in Section \ref{sec:formulation} with degree distribution pair $(\lambda,\rho)$, and parameters $a$ and $p$ heals if
\begin{equation}\label{eq:nec}
x_{0} < \frac{1}{\Big(a-1\Big)\Big( 1 + p \lambda^{\prime}(1)\Big)^{2}}
\end{equation}
\end{thm}

Theorem \ref{th:nec} provides some interesting intuitions. First, note that
\begin{equation}\label{eq:tyyloor}
\lambda^{\prime}(1) =  \sum\limits_{i\geq1} i\lambda_i \times x^{i-1} |_{x=1} = \sum\limits_{i\geq1} i\lambda_i,
\end{equation}
is the average degree of the physical nodes. Theorem \ref{th:nec} indicates the necessity of a low average degree for the physical nodes for achieving a resilient system. This is because in our model, physical nodes with higher degrees can damage more nodes. Second, this theorem suggests that the number of physical nodes under each cyber node, $a$, should be kept small. This increases the chance of healing physical nodes since a cyber node needs to have all but one measurements to heal a physical node. Finally, the theorem implies that smaller values of $p$ are desirable, which is expected.

\subsection{Optimizing for Resiliency}\label{sec:optim}
We now study design implications of the density evolution analysis of the previous section. Based on the analysis done in Theorem \ref{th:nec}, with the given network parameters $\lambda$, $\rho$, $a$, and $p$, one could evaluate the upper bound on $x_{l-1}$. Here, we refer to this value as $\epsilon_{s}$. Also, one could employ the recursive equation of (\ref{eq:f_l1}) to find the most severe disruption that can be tolerated by the network. To this end, we formulate an optimization problem with respect to network constraints. The solution to this problem would be the values of network parameters that achieve maximum resiliency against initial disturbances. We represent the maximum initial disturbance by $\epsilon_{max}$. To obtain $\epsilon_{max}$ given $a$ and $p$, an optimization problem can be set up as follows:

\begin{equation}\label{eq:opti}
\begin{aligned}
& \underset{\{\rho_i, \lambda_i, \epsilon \}}{\text{argmax}}
&& f(x_{l-1},\lambda_i,\rho_i) \\
& \text{subject to}  && x_l = f(x_{l-1},\lambda_i,\rho_i),  \\
&&& \sum_{i\geq2}\lambda_i=1, \\
&&& \sum_{i\geq2}\rho_i=1, \\
&&& 0 \leq \lambda_i \leq 1, \\
&&& 0 \leq \rho_i \leq 1. \\
\end{aligned}
\end{equation}
Sometimes, for simplicity of analysis, we assume that $\rho(x) = x^M$ for some $M\geq2$. We numerically solve this optimization problem for two scenarios. In the first case, we fix the network parameters and find $\epsilon_{max}$. In the second case, we run the program to reveal $\lambda_i$s that give us the largest $\epsilon_{max}$. We will comprehensively discuss these results in Section \ref{sec:results}.

Nevertheless, it is worth noting that the application of our proposed message-passing framework is not limited to the particular setting explained in Section \ref{sec:formulation}. This framework could be applied to any set of contagion models, healing rules, and network structures for which a density evolution analysis could be carried out. Also, this analysis holds for delay-free CPSs where cyber nodes respond immediately to failures in the physical network. However, in practical cases, cyber nodes may need a while to process the messages, collect the information, and take action. The crucial role of this time delay in the healing process will be investigated in the next section. 

\subsection{Analysis of Message Passing with Time Delays in Cyber Nodes}\label{sec:time}
We now develop the above analysis by considering processing time delay in a CPS. To this end, we employ the definition of time slots. Previously, we have assumed that each iteration of message passing can be completely done in one time slot. However, if a failure occurs for a physical node, then the corresponding cyber node usually needs a few time slots to respond to the D message. This delay would be for a number of reasons, such as recovering data from the database, collecting data from other physical nodes, gathering information from neighboring cyber nodes, etc. Therefore, one iteration would need a few time slots in order to be accomplished. 
Cyber nodes usually react against a failure in a few time slots. In what follows, we derive the density evolution equation assuming that each cyber node needs two time slots to respond to a failure. \color{black} For the purpose of simplicity and without loss of generality, we assume that messages are delivered at the destination nodes (i.e., $P_{mi} = P_{mc} = P_{mp} = 0$). \color{black} For brevity, we skip most details of the definitions and give the final equations that describe the messages at each time slot. 

\begin{thm}
For the cyber-physical system defined in section \ref{sec:formulation} with a cyber-node processing delay of two time slots, a density evolution equation for the $l$-th iteration can be obtained as 
\begin{align} \label{DE_T}
\notag x_l(t+3) &= f(x_{l-1}(t)), \\
f(x_{l-1}(t)) &= A \times B + C \times \big[1 - B \big], 
\end{align}
where A, B, and C are given as
\begin{flalign}
\notag A &= \Bigg\{ \lambda \bigg[ 1 - p \bigg( \lambda \Big(1-p x_{l-1}(t)  \Big) \times \Big(x_{l-1}(t) - 1 \Big) + 1 \bigg) \bigg] \Bigg\} \\
\notag &\times \Bigg\{ \lambda \Big(1-p x_{l-1}(t)  \Big) \times \Big(x_{l-1}(t) - 1 \Big) \Bigg\} + 1, \\
\notag B & =  \ 1 - \bigg\{ \lambda \Big(1-p x_{l-1}(t)  \Big) \times \Big(x_{l-1}(t) - 1 \Big)  \bigg\}^{(a-1)} \;  \times \\
\notag & \ \ \ \ \;  \bigg\{ 1- \rho\Big(x^a_{l-1}(t) \Big)  \bigg\}, \\
\notag C & = \ 1 - \lambda\Big( 1 - p \; A \Big). 
\end{flalign}
The system heals if and only if $x_l(t+3)\rightarrow 0$ as $l\rightarrow\infty$ (or equivalently, $t\rightarrow\infty$). 
\end{thm}
Extending the DE analysis to cyber node delays of more than two time slots can be done along the same lines as above. In the next section, through numerical results,  we will fully study the impact of different processing time delays on a self-healing cyber-physical network.

\section{Simulation Results} \label{sec:results}
To make more concrete sense of the above analysis, we have numerically simulated the message passing over cyber-physical networks. First, we simulate the network without considering processing-time delay to find the role of each network parameter. Here, we referred to this simulation as scenario I. Next, we will investigate the performance of the self-healing method in the presence of a processing-time delay. This simulation creates scenario II. Finally, we will compare both scenarios to provide clear criteria for choosing network parameters for the desired resiliency of a CPS. 

\subsection{Numerical Results for Scenario I}
This section provides a number of evaluations for the self-healing method without considering a processing-time delay. To begin with, Fig. \ref{fig:S1_0} shows the fraction of physical node failures for different numbers of iterations, $l$. As can be seen, if $l \rightarrow \infty$, then the function goes to a step function. This implies two steady-state conditions, which would be either a complete healing or a complete failure scenario in the network. These steady-state conditions confirm Theorem \ref{th:Steady_State} in Section \ref{sec:Fixed_Point_Analysis}.

\color{black}
Fig. \ref{fig:ER_SF} shows the fraction of physical nodes failure in a steady-state condition, $x_\infty$, after an initial disturbance,  for two popular network models: Erd$\ddot{\text{o}}$s-R$\acute{\text{e}}$nyi (ER) networks and scale-free (SF) networks. For this simulation, we assumed that: i) both networks have the same average degree, which is equal to $1.4$, and ii) both networks have the same minimum and maximum degrees ($k_{min}$ and $k_{max}$) that can be obtained via $k_{max} = k_{min} N ^ {(\frac{1}{\gamma - 1})}$ \cite{barabasi2000scale-free} in which $k_{min}=1$, $N = 100$, and $\gamma = 2.8$ (for SF networks). Based on these assumptions, we ran the simulations and found that SF networks have more tolerance against an initial perturbation in comparison to ER networks, as shown in Fig. \ref{fig:ER_SF}. The reason could be that, given the same average degree, most nodes in a SF network have a lower degree than the nodes in an ER network. Therefore, given that the initial failures are selected randomly and unbiased, such nodes are less likely to be among the hubs in a SF network. This means less number of high-degree failed nodes, hence, less chance of failure propagation in the network. This finding confirms the results of Schneider et al. \cite{schneider2013towards} stating that high-betweenness nodes should be planned as autonomous nodes, in order to have the best resiliency in an interdependent network. 

\begin{figure}[t]
\centering
 {\includegraphics[width =3.4 in , height=2.4 in]{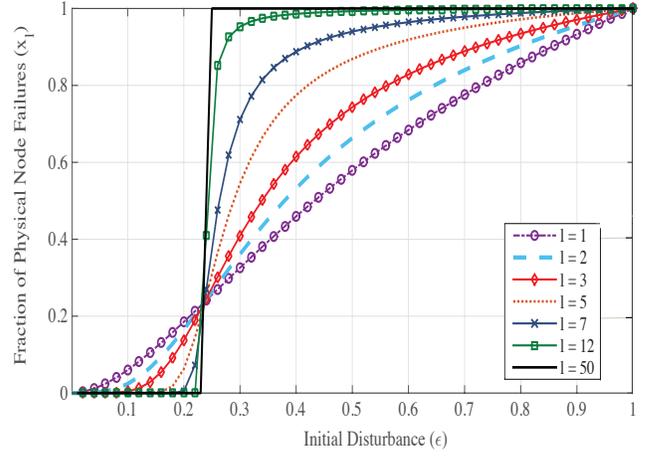}}
 \captionof{figure}{\protect\raggedright Probability of failure for physical nodes in different iterations w.r.t the initial disturbance, with network parameters $a = 5$,$\ p = 0.2$,$\lambda(z)=z ^ 2$ and  $\rho(z) = z ^ 3$.}
\end{figure}\label{fig:S1_0}
\begin{figure}[t]
\centering
{\includegraphics[width =3.5 in , height=2.4 in]{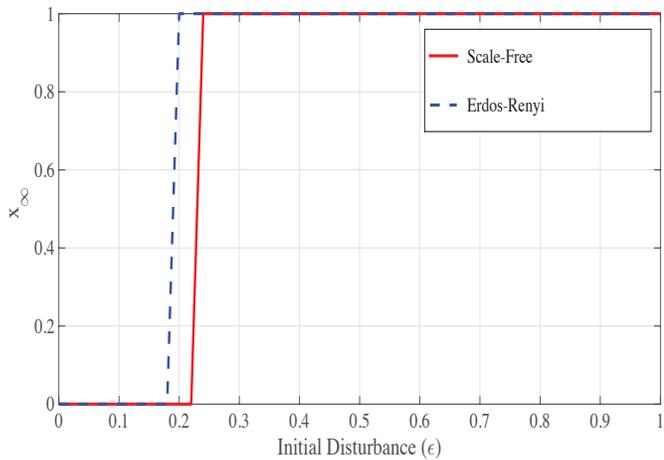}}
\captionof{figure}{\color{black} \protect Steady-state fraction of physical failed nodes against an initial disturbance for Erd$\ddot{\text{o}}$s - R$\dot{\text{e}}$nyi (ER) and Scale-Free (SF) networks. The average degree of networks is $1.4$ with the min. degree of $1$ and max. degree of $13$, and $\gamma = 2.8$.}
\end{figure}\label{fig:ER_SF}

\begin{figure}
    \centering
    \subfigure[]
    {
        \includegraphics[width =3.4 in , height=2.5 in]{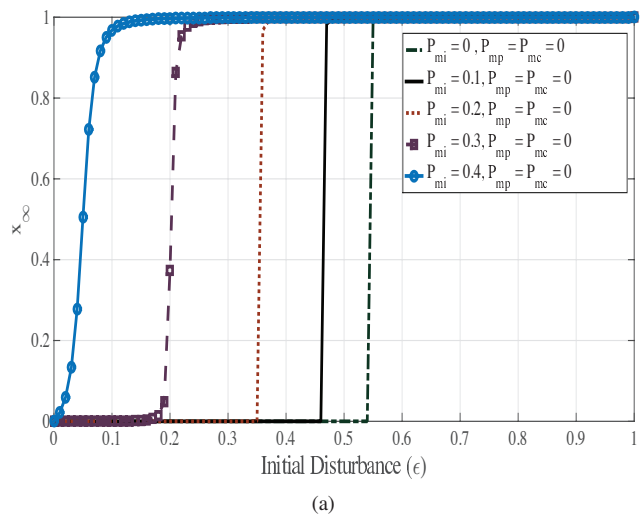}  
        \label{fig:Sm_1}
    }
    \\
    \subfigure[]
    {
       \includegraphics[width =3.4 in , height=2.5 in]{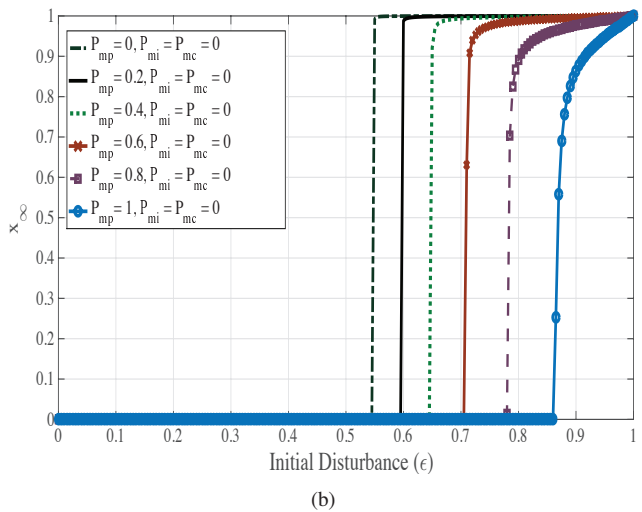}
        \label{fig:Sm_2}
    }
    \caption{Influence of missing messages on probability of nodes' failure in CPS with following parameters: $a = 4,\ p = 0.1, and \ \lambda(z) = \rho(z) = 0.5\,z + 0.4\,z ^ 2 + 0.1\,z ^ 3 $: (a) impact of missing messages between physical network and cyber network, $P_{mi}$, and (b) effect of missing messages in physical network, $P_{mp}$.}
    \label{fig:Sm}
\end{figure}

In order to study the impact of missing messages on the resiliency of a self-healing interdependent network, we simulated our findings for a CPS with different values of $P_{mi}$, $P_{mp}$, and $P_{mc}$, and the following network parameters: $a = 4,\ p = 0.1$, and $\ \lambda(z) = \rho(z) = 0.5\,z + 0.4\,z ^ 2 + 0.1\,z ^ 3$.  The results are shown in Fig. \ref{fig:Sm}. As can be seen in Fig. \ref{fig:Sm_1}, without missing messages between the two networks (i.e., $P_{mi} = 0$), the system is able to tolerate up to $55\%$ initial loss of physical nodes. However, if $P_{mi}$ gradually increases, then the opportunity of receiving $H$ messages at physical nodes from cyber nodes continually decreases. Accordingly, the resiliency of the network is drastically reduced. This trend continues until the system completely loses its ability to heal a failure ($P_{mi} \geqslant 0.4$ for the assumed network). Fig. \ref{fig:Sm_2} shows that the increase in missing messages in the physical network (i.e., $P_{mp}$) improves the probability of healing in the system. However, this result is not surprising because a larger number of missing failure messages in the physical network means a smaller chance of failure propagation in this network. In our simulations, we also noted that changes in $P_{mc}$, the probability of a missing message within the cyber network, do not affect the network resiliency significantly, since the cyber nodes in our model mainly rely on information from their own physical nodes for their operation.

\begin{figure}[t]
    \centering
    \subfigure[]
    {
       \includegraphics[width =3.4 in , height=2.5 in]{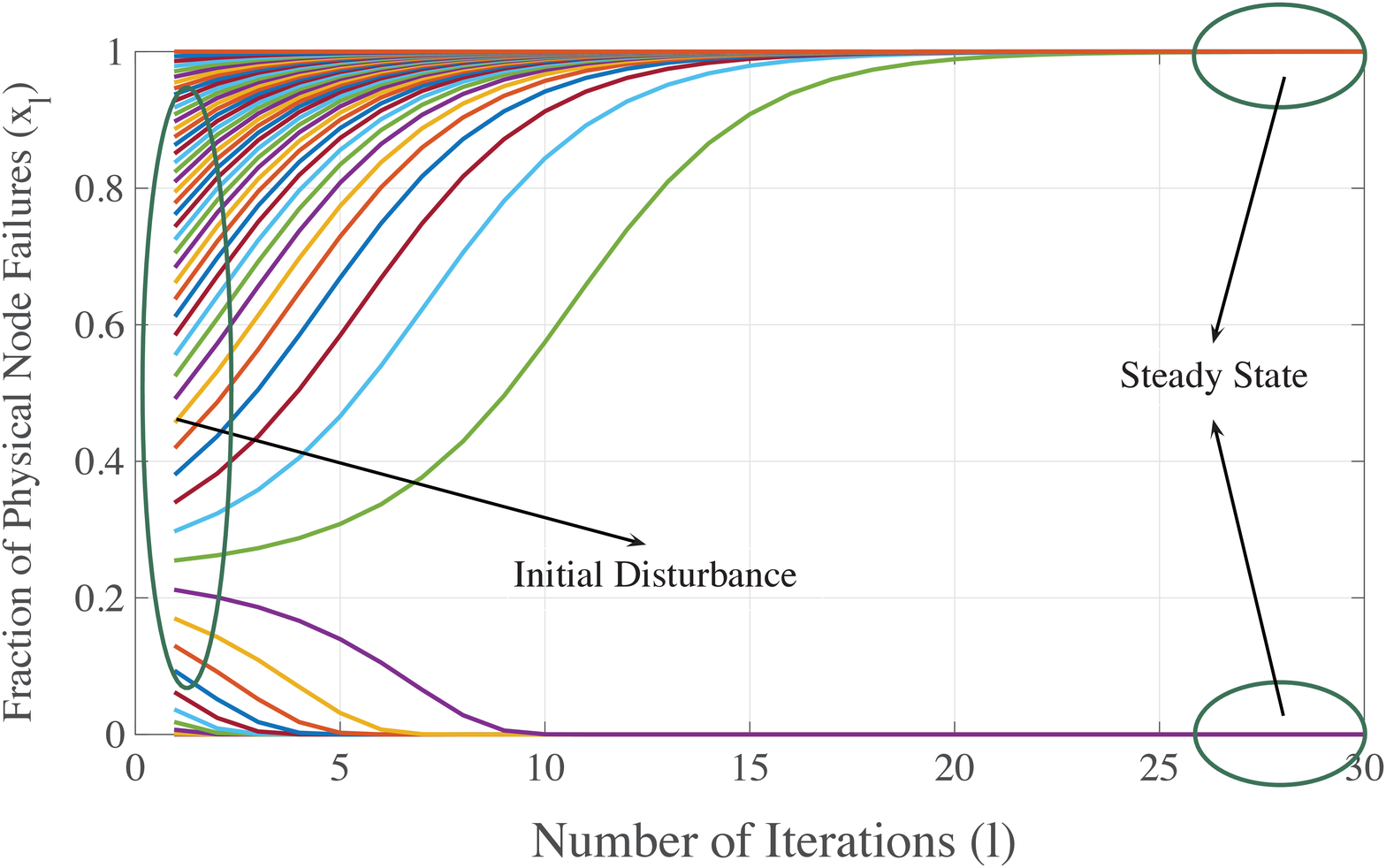}
        \label{fig:S1_a}
    }
    \\
    \subfigure[]
    {
       \includegraphics[width =3.4 in , height=2.5 in]{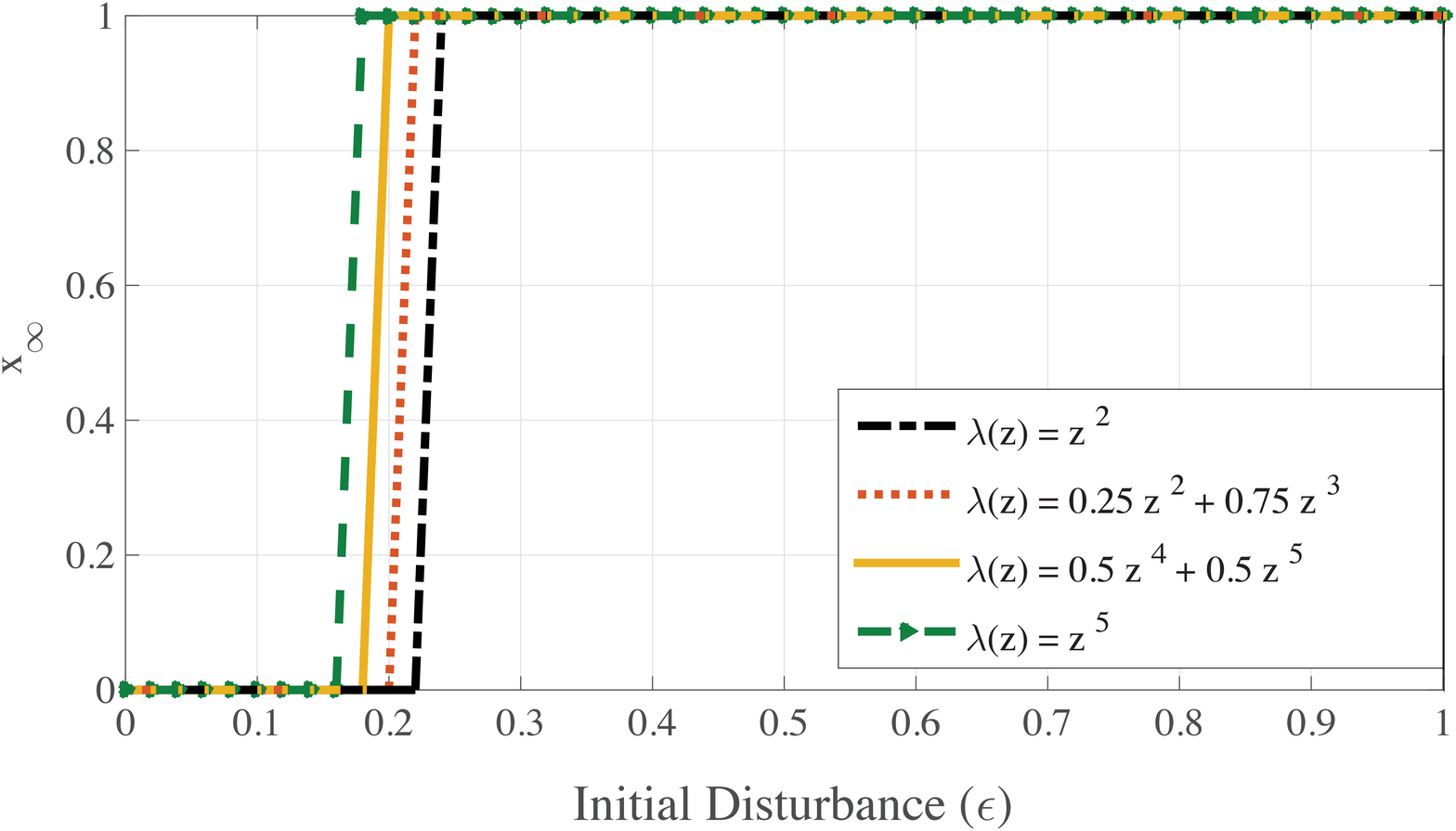}
        \label{fig:S1_b}
    }
    \caption{For network parameters $a = 5,\ p = 0.2 $ and  $\rho(z) = z ^ 3$: (a) demonstrates the number of iterations needed for a network to be completely healed or failed for $\lambda(z)=z ^3$ and (b) shows the probability of  failure for physical nodes with different $\lambda(z)$ against an initial disturbance.}
    \label{fig:S1}
\end{figure}

\color{black}Fig. \ref{fig:S1_a} indicates the number of iterations needed for the network to reach a steady-state condition. As can be seen in Fig. \ref{fig:S1_a}, the steady-state conditions would change from a complete healing (collapse) to a complete collapse (healing) for a {\em threshold} value of initial disturbance  ($\epsilon$), in this case $\epsilon=0.23$. This threshold would be varied for different sets of network parameters. The more we move away from the threshold, the less iterations are needed for the network to reach its stable state. This is due to the limited capability of cyber nodes in healing the physical nodes. For example, for small disturbance such as $ \epsilon = 0.05 $, cyber nodes need only a few iterations to heal the network because almost all physical nodes are healthy. For larger disturbances, the chance of complete healing drops rapidly. Fig. \ref{fig:S1_b} shows the threshold that can be tolerated for a set of different physical degree distributions in the network, while $a$ and $p$ are fixed. Also, we assume that enough message-passing iterations have already been done for the network to reach the steady state. As can be seen, the lower physical degree distributions have a higher resiliency against failure.

In order to find the maximum threshold and network parameters achieving this threshold, we numerically solve the optimization problem of (\ref{eq:opti}).
The problem is solved for two cases: (a) $a$, $\lambda(\cdot)$, $\rho(\cdot)$, and $p$ are kept fixed and the maximum tolerated initial disturbance, $\epsilon_{max}$, is found, and (b) $a$, $p$, and $\rho(\cdot)$ are fixed while $\lambda_i$s are obtained for the $\epsilon_{max}$. Results for the first and the second cases are shown in Table I and II, respectively, where the corresponding value of $\epsilon_{s}$, the value of the upper bound in (\ref{eq:nec}), is also listed for each set of network parameters. 
The following observations can be made from Table I:
\begin{itemize}
\item The results in Table I.A indicate that increasing $a$ reduces the values of $\epsilon_{s}$ and $\epsilon_{max}$. In fact, a large $a$ increases the chance of receiving D messages by a cyber node from its physical neighbors. This reduces the resiliency of the system.
  \item The results in Table I.B confirm that reducing $p$ leads to less vulnerability of physical nodes from their physical neighbors.
  \item Table I.C shows that a less-connected physical network results in larger values of $\epsilon_{s}$ and $\epsilon_{max}$ and, hence, higher resiliency.
\end{itemize}

\begin{table}[t]
\centering
\caption{$\epsilon_{s}$ and $\epsilon_{max}$ for different network parameters and severity of initial disturbance}
\label{tab:f(x0)}
 \begin{tabular}{ | c | c | c | c || c | c | }
                \multicolumn{6}{c}{}\\
  \multicolumn{6}{c}{Table I.A: $\epsilon_{s}$ and $\epsilon_{max}$ for variation of $a$}\\
   \hline
   $p$ & $a$ & $\lambda(z)$ & $\rho(z)$ & \color{red}{$\epsilon_{s}$} & \color{red}{$\epsilon_{max}$} \\ \hline
   \multirow{3}{*}{0.8} & $3$ & \multirow{3}{*}{$z^{2}$} &  \multirow{3}{*}{$z^{3}$} & \color{red}{0.0740} & \color{red}{0.1002} \\
   & $5$ & & & \color{red}{0.0369} & \color{red}{0.0482} \\
   & $8$ & & & \color{red}{0.0211} & \color{red}{0.0271} \\ \hline
               \multicolumn{6}{c}{}\\
  \multicolumn{6}{c}{Table I.B: $\epsilon_{s}$ and $\epsilon_{max}$ for variation of $p$}\\
   \hline
   $p$ & $a$ & $\lambda(z)$ & $\rho(z)$ & \color{red}{$\epsilon_{s}$} & \color{red}{$\epsilon_{max}$} \\ \hline
    0.4 & $\multirow{3}{*}{4}$ & \multirow{3}{*}{$z^{2}$} &  \multirow{3}{*}{$z^{3}$} & \color{red}{0.1028} & \color{red}{0.1621} \\
   $0.6$ & & & & \color{red}{0.0688} & \color{red}{0.0973} \\
   $0.8$ & & & & \color{red}{0.0493} & \color{red}{0.0650} \\ \hline
                \multicolumn{6}{c}{}\\
  \multicolumn{6}{c}{Table I.C: $\epsilon_{s}$ and $\epsilon_{max}$ for variation of $\lambda(x)$}\\
   \hline
   $p$ & $a$ & $\lambda(z)$ & $\rho(z)$ & \color{red}{$\epsilon_{s}$} & \color{red}{$\epsilon_{max}$} \\ \hline
    $\multirow{3}{*}{0.5}$ & $\multirow{3}{*}{3}$ & $z^{2}$ &  \multirow{3}{*}{$z^{3}$} & \color{red}{0.1250} & \color{red}{0.1933} \\
    & & $z^5$ & & \color{red}{0.0408} & \color{red}{0.0525} \\
    & & $z^8$ & & \color{red}{0.0200} & \color{red}{0.0250} \\ \hline
 \end{tabular}
\end{table}

\begin{table}[t]
\centering
\caption{$\epsilon_{s}$ and $\epsilon_{max}$ for different degree coefficients of physical nodes.}
\label{tab:f(x0,lambda)}
 \begin{tabular}{ | c | c | c || c | c | c | }
\hline
   $p$ & $a$ & $\rho(z)$ & \color{red}{$\lambda(z)$} & \color{red}{$\epsilon_{s}$} & \color{red}{$\epsilon_{max}$} \\ \hline
   \multirow{4}{*}{0.5} & \multirow{4}{*}{$4$} & \multirow{4}{*}{$z^{3}$} & \color{red}{$\lambda_2 = 1$} &\multirow{4}{*}{\color{red}{0.05334}} & \multirow{4}{*}{\color{red}{0.07245}} \\
   & & & \color{red}{$\lambda_3 = 0$} & & \\
   & & & \color{red}{$\lambda_4 = 0$} & & \\
   & & & \color{red}{$\lambda_5 = 0$} & & \\ \hline
   \multirow{4}{*}{0.2} & \multirow{4}{*}{$3$} & \multirow{4}{*}{$z^{2}$} & \color{red}{$\lambda_2 = 1$} &\multirow{4}{*}{\color{red}{0.19531}} & \multirow{4}{*}{\color{red}{0.35424}} \\
   & & & \color{red}{$\lambda_3 = 0$} & & \\
   & & & \color{red}{$\lambda_4 = 0$} & & \\
   & & & \color{red}{$\lambda_5 = 0$} & & \\ \hline
 \end{tabular}
\end{table}

In the second case, we consider a physical degree distribution of minimum degree two and maximum degree five:
\begin{equation}\label{eq:lambda_sim}
\lambda(z) = \lambda_2 \: z^2 \; + \; \lambda_3 \: z^3 \; + \; \lambda_4 \: z^4 + \; \lambda_5 \: z^5.
\end{equation}
The outcomes for this case are shown in Table \ref{tab:f(x0,lambda)}. As can be seen, we consistently obtain $\lambda_2 = 1$, while $\lambda_3 = 0$, $\lambda_4 = 0$, and $\lambda_5 = 0$. This is not surprising, because physical nodes with more physical neighbors help spread the failures. In fact, when solving the optimization equation (\ref{eq:opti}) for scenarios with other limitations on $\lambda(z)$, we found that the degree of the physical nodes should be kept as small as possible.

\vspace{-.04 in}
\subsection{Numerical Results for Scenario II}
\begin{figure}[t]
    \centering
    \subfigure[]
    {
       \includegraphics[width =3.4 in , height=2.2 in]{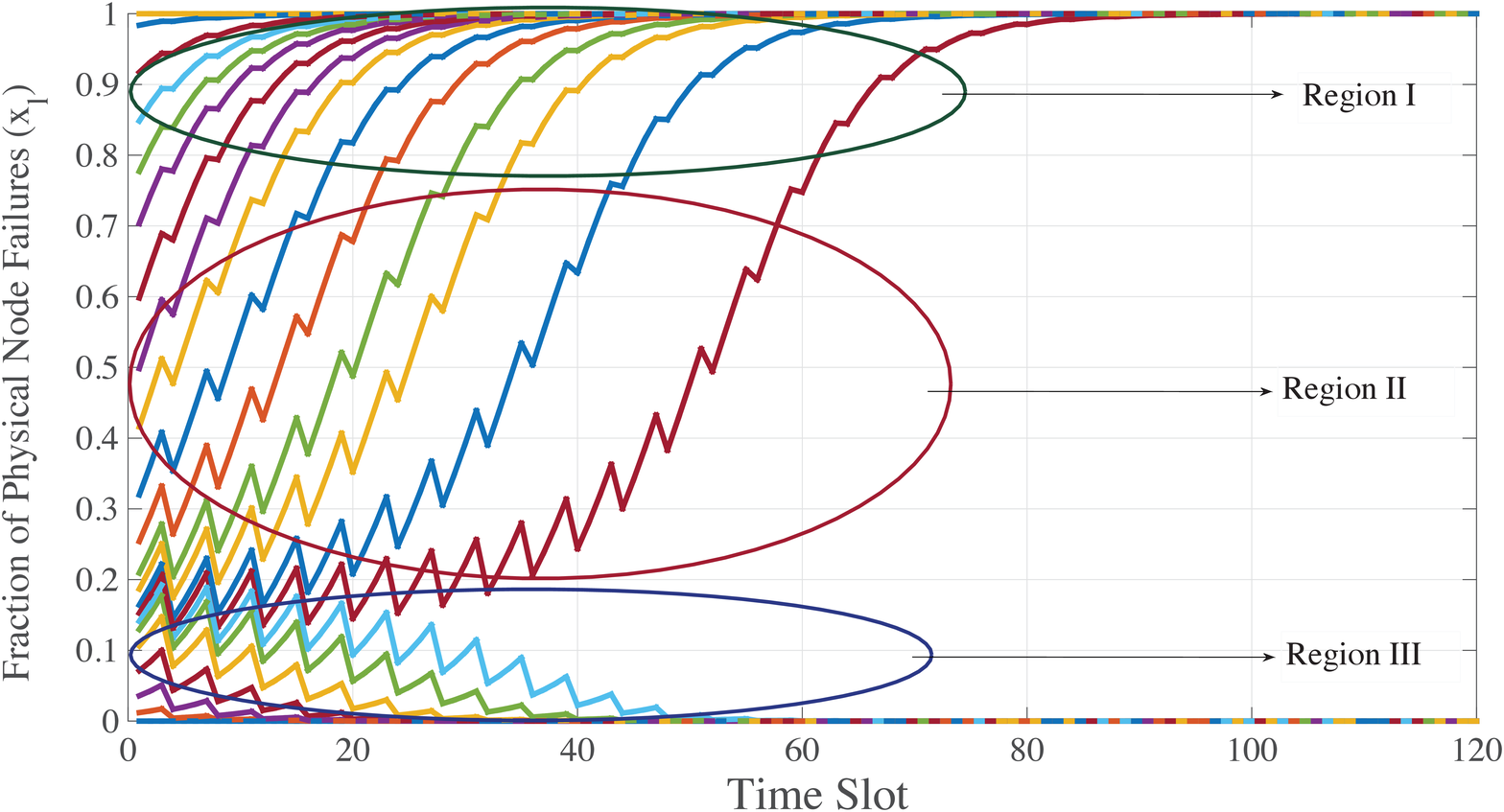}
        \label{fig:S2_a}
    }
    \\
    \subfigure[]
    {
       \includegraphics[width =3.4 in , height=2.2 in]{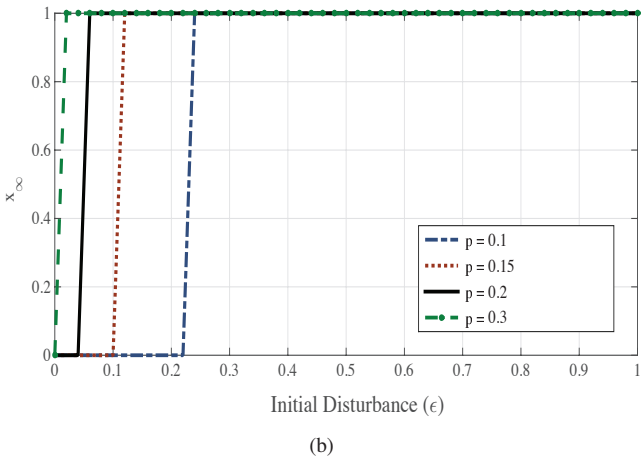}
        \label{fig:S2_b}
    }
\caption{(a) The probability of failure for physical nodes in presence of processing-time delay. Processing delay = 3 time slots, and network parameters are $a = 5,\ p = 0.2, \lambda(z) = z ^ 2$, and $\rho(z) = z ^ 3$. (b) Impact of $p$ on steady-state behavior of network.} \label{fig:S2}
\end{figure}

In Section \ref{sec:time}, we have analytically obtained the influence of processing-time delay on the probability of failure in the CPS. Now, we numerically evaluate those results in the same line of scenario I. We begin with a processing-delay of three time slots. Hence, each iteration can be accomplished in four time slots. Fig. \ref{fig:S2_a} shows the number of time slots needed for the network to reach a steady-state condition. As can be seen, jagged lines occur as a result of the processing-time delay. Considering this delay, a cyber node needs $k$ time slots (here, $k = 3$) to respond to the failure.
During this time interval,  the failure propagates in the physical network and could constantly increase the probability of failure for physical nodes. After the $k$-th time slot, however, the network would experience three cases for the probability of failure, depending on the ability of cyber nodes to heal their associated physical nodes: 
\begin{itemize}
\item Region I: The healing ability of cyber nodes is diminishing due to the large number of failures in the network; hence, the probability of physical failures only increases. 
\item Region II: Cyber nodes are still able to heal their physical nodes, but this is not enough to stop the propagation of failure and change the overall trend. That is, the number of failures at the end of the iteration is still higher than that of the beginning of the iteration.
\item Region III: The healing ability of cyber nodes outweighs the propagation of failure, leading to complete healing of the network after a few iterations. 
\end{itemize}

The probability that each physical node gets affected by the failure of its physical neighbors, $p$, is decisive to the resiliency of a CPS. One example of this parameter in power systems could be revealed in protective relays. The mission of protective relays is to sense a fault and initiate a trip, disconnection or order. Therefore, good relays result in lower probability of failure in a network. The performance of relays, in a very abstract sense, can be mapped to $p$. The influence of $p$ becomes more crucial, when there is a processing-time delay for cyber nodes. Fig. \ref{fig:S2_b} shows the steady-state behavior of the network for different values of $p$, when the processing-time slot is three ($k=3$). As can be seen, the higher value of $p$ dramatically increases the vulnerability of the network against an initial disturbance. For instance, for $p=0.3$, network resiliency is almost non-existent against any initial disturbance in the physical network.

\subsection{Comparison of Results for Scenario I and II}

Now, we compare scenarios I and II to understand the effect of delay on the performance of cyber-physical systems. First, consider Fig. 12(a), which shows the direct impact of delay on maximum resiliency of a network. Delayed systems need considerably more time slots to be healed in comparison to non-delayed systems. This can be observed for $\epsilon = 0.1$. In addition, at $\epsilon = 0.3$, the non-delayed system can be cured after nine time slots. However, the delayed system completely collapses. The reason is that the failure propagates throughout the physical network during the three time slot delay interval. Fig. 12(b) displays the steady-state behavior for delayed and non-delayed systems w.r.t. the size of the initial disturbance. A comparison of resiliency thresholds effectively demonstrates the essential need for a quick response from cyber nodes.

\begin{figure}[t]
    \centering
    \subfigure[]
    {
       \includegraphics[width =3.4 in , height=2.2 in]{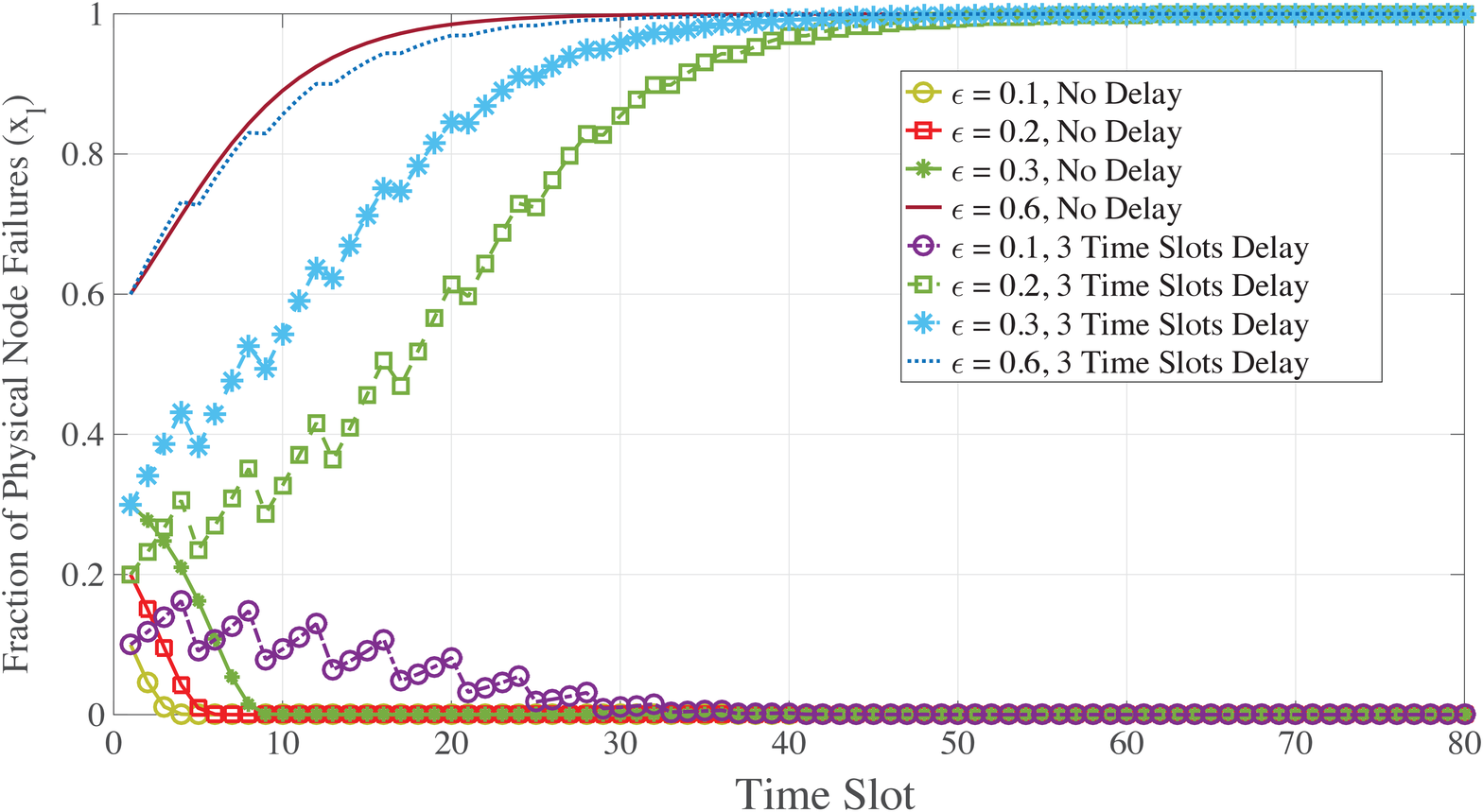}
        \label{fig:S3_a}
    }
    \\
    \subfigure[]
    {
      \includegraphics[width =3.4 in , height=2.2 in]{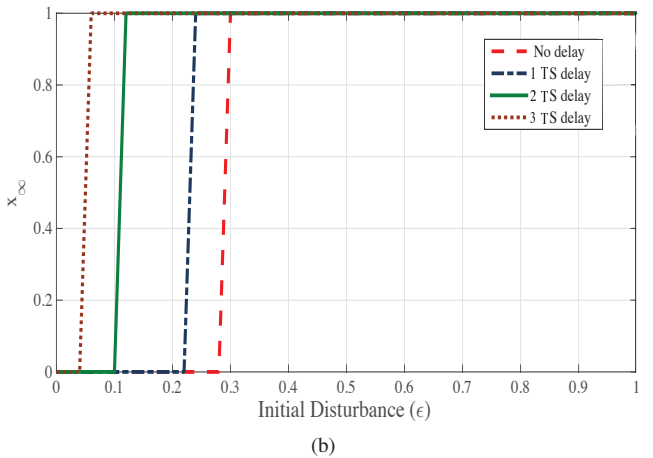}
        \label{fig:S3_b}
    }
\caption{Comparison between non-delayed systems and delayed systems for network parameters  $a = 5,\ p = 0.15, \lambda(z) = z ^ 2$, and $\rho(z) = z ^ 3$: (a) effect of time slot processing delay on systems, and (b) impact of time slot (TS) delays on maximum tolerated threshold in network.} 
\end{figure}\label{fig:S3}

\section{Conclusion}\label{sec:conclusion}
We introduced a graphical model representation of cyber-physical systems and applied message passing to investigate the resiliency of inter-dependent CPSs. We provided a density evolution analysis to study the behavior of a system in the presence of both self-healing and propagation of failures. Our analysis resulted in a sufficient condition on choosing the network parameters for the system to completely heal after an initial disturbance. Then, we studied the steady-state behavior of cyber-physical networks after an initial disturbance, proving that the network reaches one of the two conditions, either a complete healing or a complete failure. To improve the network robustness, we set up an optimization problem to calculate network parameters for highest network resiliency against physical node disruptions, where we found the most severe attack that can be tolerated by the network, given a set of parameters. Finally, we studied self-healing cyber-physical networks where the response of cyber nodes to failures in the physical network is delayed. Our findings revealed the crucial importance of low processing-time delay for increasing the probability of healing in the network. 

\section{ Appendix}
\color{black}
\subsection{Proof of Theorem I}
\begin{proof}
After a disturbance has occurred, D messages will be generated by the failed nodes. Since in our model each node is disturbed by a probability $\epsilon$, the initial density of D's, denoted by $x_0$, is $\epsilon$. Assume that we are at the beginning of the $l$-th iteration and find $x_l$ in terms of $x_{l-1}$. In order to obtain the recursive equation during the $l-1$-th iteration, we need to define $y$, $u$, and $w$ rigorously. To this end, we suppose that exchanged messages are independent of each other. The next theorem shows that this assumption holds if the number of nodes is sufficiently large in the network. According to the definition of failure messages from a physical node to a cyber node, and the probability of missing messages, $y$ can be written as (22). Also, the failure message from a cyber node to a physical node, $u$, can be derived as (23). 
\begin{figure*}[b]
\begin{align}
\toprule
y= \; & \text{Pr} \: \Bigg\{ \bigg(\text{Physical node itself has failed}\bigg) \quad \bigcup \quad \bigg( \text{Node has not failed} \;  \bigcap \; \text{At least one of its}
\notag \\
& \: \quad \; \text{physical neighbors has failed} \quad  \bigcap \quad \text{Failed message from the neighbor has not missed}  \bigg) \Bigg\} \notag  \\
 \Rightarrow y = \; & x_{l-1} + \Big(1-x_{l-1}\Big)\Big(1 - \lambda(1- p x_{l-1})\Big) \Big( 1- P_{mp} \Big) \label{eq:x1_v}.
\end{align} 
\begin{align}
u & = \text{Pr} \: \bigg\{\text{Cyber node received the failure of at least one of its physical nodes} \:\bigcap \: 
\notag \\
& \: \quad \quad \quad \; \text{Cyber node has access to at least one healthy cyber node}\Big\}, \notag \\
\Rightarrow  u & = 1 - \bigg(\Big({1 - y}\Big) \Big( 1 - P_{mi}\Big) \bigg)^{a-1}\bigg(1 - \rho\Big(w \, (1-P_{mc}) \Big)\bigg) \label{eq:x2_v}.
\end{align} 
\begin{align}
x_{l} &= \text{Pr} \: \Bigg\{\text{\bigg(Physical node fails} \ \bigcap \ \text{Its cyber node sends D message} \bigg) \ \ \bigcup  
\notag \\
&\quad \qquad \; \text{\bigg(Physical node fails} \ \bigcap \ \text{H message from its cyber node is missed} \bigg) \Bigg\} \notag \\
\Rightarrow x_{l} & = y \times u \ + \ y \times (1 - u)\, P_{mi} 
\end{align} 
\end{figure*}
Finally, the failure message between cyber nodes, $w$, is defined as
\begin{align}
w &= \text{Pr} \bigg\{\text{Cyber node receives message of failure of all}  \notag \\
&  \quad \text{of its physical nodes}\bigg\} 
 \Rightarrow w = \Big(y \, \big(1 - P_{mi} \big) \Big)^{a}.\label{eq:x3_v}
\end{align}
At the $l$-th iteration, $x_{l}$ can be defined as (24). We can eliminate $w$ by substituting (25) into (23) and obtaining $u$ as a function of $y$. Substituting (22) into (23) and then (24) yields (\ref{eq:f_l1}).
\end{proof}

\vspace{- 0.05 in}

\vspace{-0.1 in}
\subsection{Proof of Theorem II}

\begin{figure}[t]
\centering
{\includegraphics[width =3.0 in , height=1.8 in]{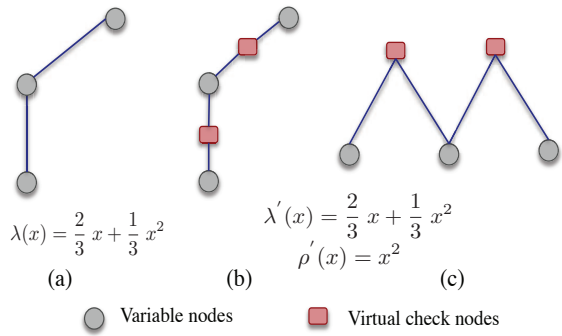}}
\caption{Example of bipartite graph with virtual check nodes: (a) variable nodes, (b) virtual check node inserted between every two variable nodes, and (c) variable nodes and virtual check nodes forming a bipartite graph.}\label{fig:Cycle-free}
\end{figure}

\begin{proof}
Let us divide all messages in a CPS into two parts: (a) messages between two networks, or inter-messages, and (b) messages within networks, or intra-messages. The edges and nodes for the inter-messages create the form of bipartite graphs. Richardson and Urbanke \cite{richardson2001capacity} showed that if the length of the smallest {\em cycle} in this bipartite graph is greater than $2l$, or in other words, if the neighborhood of length $2l$ for every node is {\em cycle-free}, then, up to the $l$th iteration, the incoming messages to each check node or variable node are independent from each other. The authors then proved in Appendix A of \cite{richardson2001capacity} that, for any given $l$, such a cycle-free structure is achieved for sufficiently large number of nodes. The same exact proof applies to the bipartite graph connecting the cyber and physical networks as the number of nodes grows very large, thereby guaranteeing the independence of {\em inter}-messages.

There are no intra-messages in the message passing in codes on graphs as the Tanner graph is bipartite. The existence of intra-messages in CPSs is a direct consequence of the third difference listed in Section \ref{sec:BP} between the codes and CPSs.
To obtain the independence between intra-messages, we first map each node (physical or cyber) to a variable node as shown in Fig. \ref{fig:Cycle-free}(a). Without loss of generality, we add a virtual check node between every two connected variable nodes as shown in Fig. \ref{fig:Cycle-free}(b). This virtual check node acts as a relay for the messages and does not have any impact on the message passing. The resulting graph can now be simply considered as a bipartite graph as shown in Fig. \ref{fig:Cycle-free}(c). As seen in the figure, the degree distribution of variable nodes remains the same while the degree distribution of virtual check nodes always equals to $x^2$. Obtaining such a bipartite graph enables us to apply the same logic of inter-messages using \cite{richardson2001capacity} for the intra-messages of a network. That is, if the number of nodes in a network (either physical or cyber) with a given degree distribution is sufficiently large, then the network satisfies the cycle-free condition as stated above, which guarantees the independence of {\em intra}-messages. 
\end{proof}

\begin{figure}[t]
\centering
{\includegraphics[width =3.4 in , height=1.3 in]{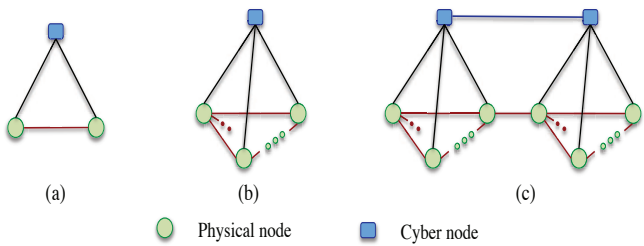}}
\caption{Illustration of cyber-physical system graph in proof of Theorem \ref{th:Steady_State}: (a) simple CPS, (b) CPS with one cyber and $n$ physical nodes, and (c) CPS with two cyber nodes and $n$ physical nodes.}\label{fig:Steady_State}
\end{figure}

\vspace{-0.1 in}
\subsection{Proof of Theorem III}
\begin{proof}
We first show that the theorem holds for a simple network. We then extend the theorem truth to more general networks. To begin with, consider a network with one cyber node and two physical nodes, as shown in Fig. \ref{fig:Steady_State}(a). If $\mu_1$ denotes the probability of failure for a physical node, then one of the following probabilities may occur:
\[   \left\{
\begin{array}{ll}
      \dbinom{2}{0} \ \mu_1^ 0 \  \big(1 - \mu_1\big)^2         &   \quad    \quad   Case \ I: 0 \ node \ failure,                                      \\
      \\
      \dbinom{2}{1} \ \mu_1 \ \big(1 - \mu_1\big)                &   \quad      \quad  Case \ II: 1 \ node \ failure,                                        \\
      \\
      \dbinom{2}{2} \ \mu_1^2 \ \big(1 - \mu_1\big)^0        &   \quad     \quad   Case \ III: 2 \ nodes \ failure.                                     \\
\end{array} 
\right. \]
For cases I and III, the system is in a steady-state condition. In case I, none of the nodes is affected by the failure and the network is healthy. In case III, two nodes are lost due to failure. As we assumed (like LDPC codes), if more than one physical node is lost, then the corresponding cyber node cannot heal them. Hence, the physical nodes remain failed and in turn cause the cyber node to fail. Therefore, the network goes into complete-collapse. In case II, however, the system would be in a transient condition, which means that the network has neither completely healed nor completely failed. This occurs when a 
cyber node heals the failed node, but the failed node already propagates the failure to one of its neighbors. Hence, there is a failed node in the next state. After the $l$-th iteration, the probability of the network to be in a transient condition is
\begin{align}
\dbinom{2}{1} \ \mu_1 \ \big(1 - \mu_1\big) \ \Big( (1 - p)^{(\alpha-1)} \ p \Big)^l,
\end{align}
where $p$ represents the probability of failure propagation between physical nodes, and $\alpha$ shows the number of neighbors for the failed node. For simplicity, we have assumed that all nodes have the same number of neighbors. As the $l$ grows, the probability of a network to be in a transient condition goes to zero. So, the system reaches a steady-state healing condition (case I + no failure propagation during message passing in case II) or a steady-state collapsed condition (case III + at least two failed nodes due to failure propagation in case II).
\vspace{-0.01 in}

The assumption of two physical nodes can be extended to $n$ physical nodes, as shown in Fig. \ref{fig:Steady_State}(b). In the same fashion as above, the probability of transient condition in such networks after the $l$-th iteration would be 
\begin{align}
\dbinom{n}{1} \ \mu_1 \ \big(1 - \mu_1\big)^{(n-1)} \ \Big( (1 - p)^{(\alpha-1)} \ p \Big)^l.
\end{align}
As $l \rightarrow \infty$, the probability of transient condition goes to zero.

We then increase one cyber node to $m$ cyber nodes by defining clusters. A cluster includes a cyber node and its supporting physical nodes. Fig. \ref{fig:Steady_State}(c) shows two clusters ($m=2$). By employing the above conclusion, the probability of one cluster with $k$ physical nodes being in a transient condition is
\begin{align}
\dbinom{k}{1} \ \mu_2 \ \big(1 - \mu_2\big)^{(k-1)} \ \Big( (1 - p)^{(\alpha-1)} \ p \Big)^l,
\end{align}
where $\mu_2$ is the probability of failure for a physical node in one cluster. Hence, for $v$ out of $m$ clusters with one failed physical node, the probability of a transient condition in entire network after the $l$-th iteration would be  

\begin{align}
 &\dbinom{k_1}{1} \ \mu_2 \ \big(1 - \mu_2\big)^{(k-1)} \Big( (1 - p)^{(\alpha-1)} \ p \Big)^l, \ \times \ . \ . \ . \:   \times \notag \\
  &\dbinom{k_v}{1} \ \mu_2 \ \big(1 - \mu_2\big)^{(k_v-1)}   \ \Big( (1 - p)^{(\alpha-1)} \ p \Big)^l,
\end{align}
where $k_i$ represents the number of nodes in the $i$-th cluster. As the number of iterations grows, the transient condition gradually vanishes. In other words, if during one of these iterations all nodes becomes healthy, then the network become healthy. Also, if two nodes in one cluster fails during the iterations, then the failures gradually permeate among the nodes in the cluster and then all nodes in the network. Therefore, a cyber-physical system with $m$ cyber nodes and $n$ physical nodes reaches a steady-state condition.
\end{proof}
\subsection{Proof of Theorem IV}
\begin{proof}
By taking the Taylor series from the right side of (\ref{eq:f_l1}) at $x_{l-1} = 0$, we obtain
\begin{align}\label{eq:taylor}
\notag x_{l} = \ & \Big(a-1\Big)\Big(1+p\lambda^{\prime}(1)\Big)^{2}\;x_{l-1}^{2} - 0.5\Big(a-1\Big)\Big(1+p\lambda^{\prime}(1)\Big) \ \\
\times &  \bigg[\Big(a-2\Big)\Big(1+p\lambda^{\prime}(1)\Big)+2p\Big(2\lambda^{\prime}(1)+p\lambda^{\prime\prime}(1)\Big)\bigg]\; x_{l-1}^3 \notag \\
\; + &  \; O(x_{l-1}^4).
\end{align}
For $x_l$ to be less than $x_{l-1}$, it is enough to show that $x_{l-1}$ is larger than the first term on the right side of (\ref{eq:taylor}). That is,
\begin{align}\label{eq:first_term}
\Big(a-1\Big)\Big(1+p\lambda^{\prime}(1)\Big)^{2}\;x_{l-1}^{2} < x_{l-1}.
\end{align}
Inequality (\ref{eq:first_term}) holds for every $l$ if it holds for $l=1$. Substituting $l=1$ in (\ref{eq:first_term}) leads to (\ref{eq:nec}).
\end{proof}
\subsection{Proof of Theorem V}
\begin{proof}
Let $x(t), y(t), u(t)$ and $w(t)$ denote the messages in the $t$-th time slot (see Fig. \ref{fig:Message}). If the probability of failure at the beginning of the $t$-th time slot is $\epsilon$, then we have $x(t) = \epsilon$. In the next time slot, according to the model described in Sections \ref{sec:setting} and \ref{sec:Message_Types}, we have
\begin{align}\label{t+1}
\notag y(t+1) & = \ x(t) + \Big[1-x(t)\Big] \ \Big[1-\lambda\Big(1- p x(t)\Big)\Big], \\
\notag  u(t+1) &  =\ 1 - \Big[1 - y(t)\Big]^{(a-1)} \Big[1 - \rho\Big(w(t)\Big)\Big], \\
  w(t+1) &  =\ \Big(y(t)\Big)^{a}, \quad
 x(t+1) =\ y(t+1),
\end{align}
where $y(t) = x(t)$, and $w(t) =\ \big(x(t)\big)^{a} = \epsilon \: ^a$. Similarly, in the $(t+2)$-th time slot, we obtain
\begin{align}\label{t+2}
\notag y(t+2)  = \ & x(t+1) + \Big[1-x(t+1)\Big]  \ \times \\
&  \Big[1-\lambda\Big(1- p x(t+1)\Big)\Big], \notag  \\
\notag u(t+2)  =  \ &  1 - \Big[1 - y(t+1)\Big]^{(a-1)} \Big[1 - \rho\Big(w(t+1)\Big)\Big], \\
 w(t+2)  = \ &   \Big(y(t+1)\Big)^{a}, \quad
 x(t+2)  =  y(t+2).
\end{align}
After two time slots for processing the data in a cyber node, the probability of failure of a physical node can be calculated as
\begin{align}\label{t+3}
x_l(t+3)   =  & \ y(t+2) \ u(t+2) \ + \notag \\
&  \Big[1-\lambda\Big(1- p x(t+2)\Big)\Big] \Big[1 - u(t+2)\Big].
\end{align}
If we substitute equation (\ref{t+1}) into (\ref{t+2}) and then the result into (\ref{t+3}), in the same line of proof Theorem I, equation (\ref{DE_T}) will be obtained. 
\end{proof}

\linespread{1.14}
\bibliographystyle{ieeetran}
\bibliography{biblio1}

\begin{IEEEbiography}[{\includegraphics[width=1in, height=1.35in,clip, keepaspectratio]{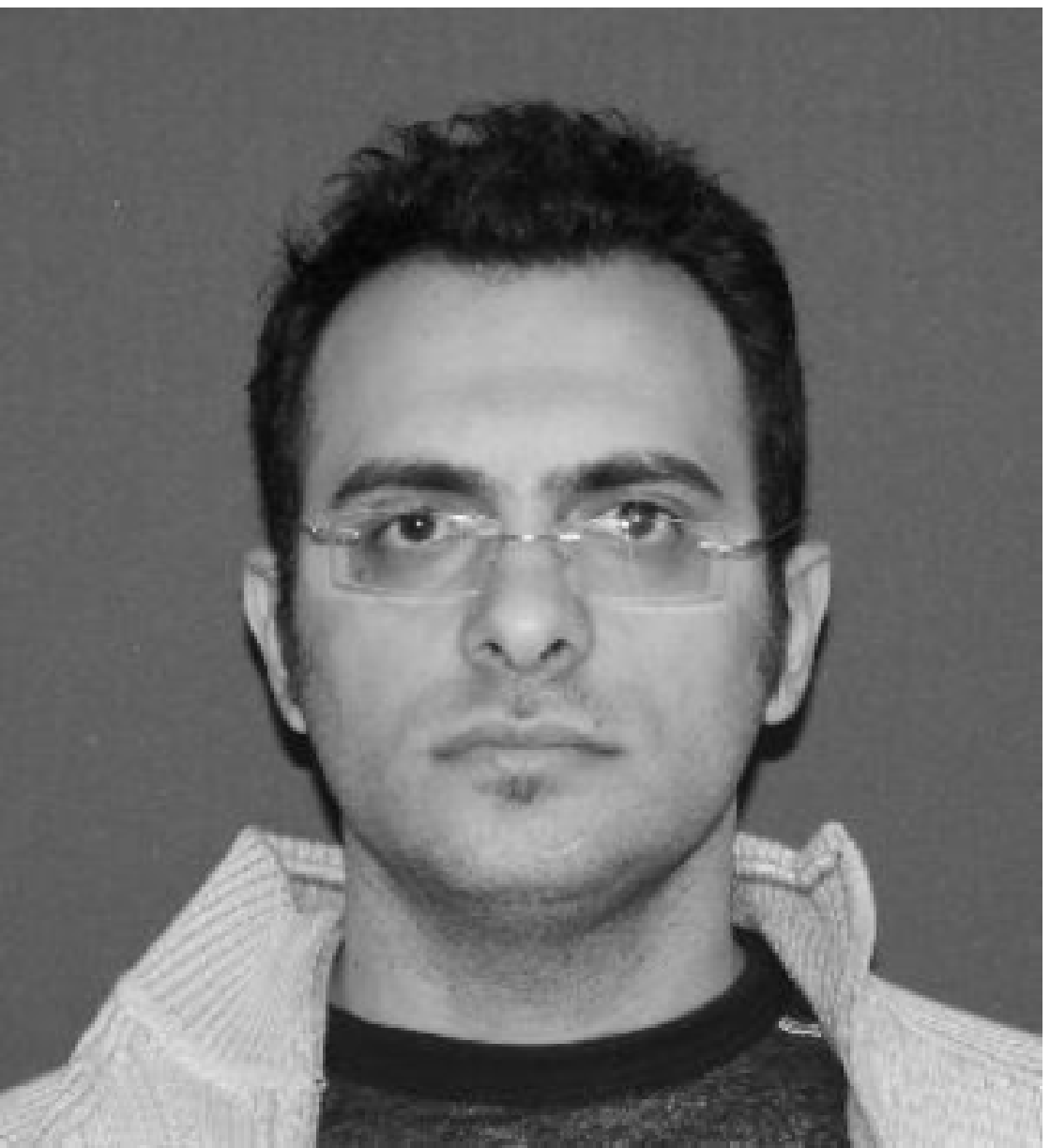}}] {Ali Behfarnia} (S'16) is a Ph.D. candidate in the Department of Electrical Engineering and Computer Science at Wichita State University, Kansas, USA. He is a recipient of the {\em GlobComm} travel grant, the Bright Future Award from Wichita State Ventures, and the Donald D. Sbarra Endowed Fellowship, all in 2016. His research interests include resilient cyber-physical system, error control coding, and communications over wireless networks. 
\end{IEEEbiography}

\begin{IEEEbiography}[{\includegraphics[width=1in, height=1.35in,clip, keepaspectratio]{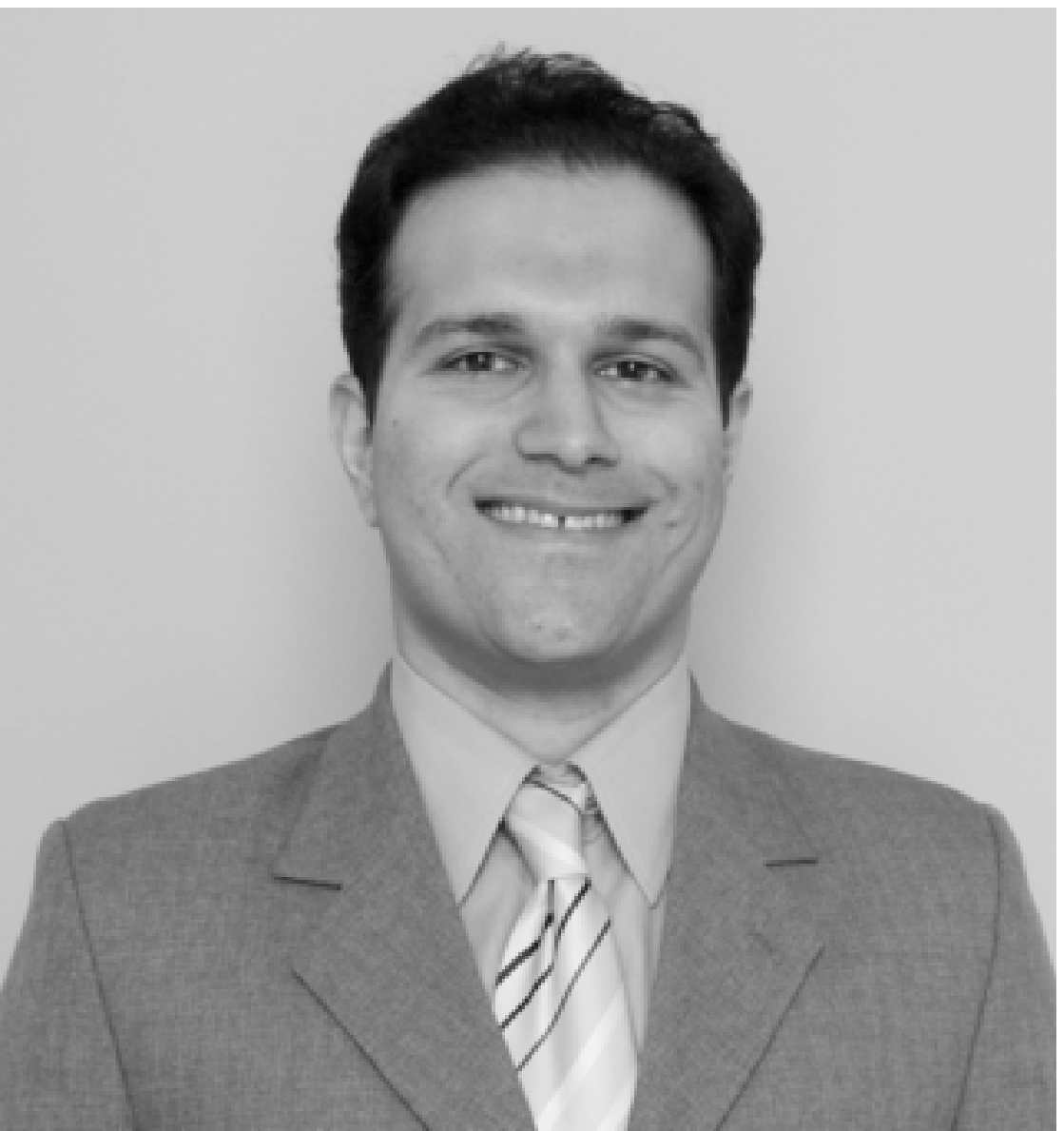}}] {Ali Eslami}
(S'07-M'13) received the PhD in electrical and computer engineering from the University of Massachusetts, Amherst, in 2013. He is an assistant professor of electrical engineering and computer science at Wichita State University, Wichita, Kansas. He is the recipient of Wichita State's Young Faculty Risk Taker Award for 2016-2017 academic year.
From August 2014 to June 2015, he worked as a visiting research scholar of information initiative at Duke (iiD). He held position as postdoctoral research fellow with Texas A\&M University, College Station, Texas, from March 2013 to April 2015. He is a member of the IEEE Communications and Computer Societies, and has served as a reviewer for numerous IEEE journals. He has also served on several NSF review panels, and as session chair in several IEEE conferences and workshops. His current research interests include nano-communications, applications of coding theory in biology, resilient design of cyber-physical systems, fault-tolerant quantum computing, and big-data storage systems. 
\end{IEEEbiography}

\IEEEpeerreviewmaketitle

\end{document}